\documentclass{emulateapj}
\usepackage{psframe}
\usepackage{lscape}
\usepackage{amsmath}
\numberwithin{equation}{section}

\begin{document}


\title{ Correcting for the Ionosphere in the $uv$-Plane }
\author{ Michael S. Matejek} \email{mmatejek@mit.edu} \affiliation{MIT Kavli Institute for Astrophysics and Space Research \linebreak
77 Massachusetts Avenue 37-287 \linebreak Cambridge, MA 02139  } \nobreak \author{Miguel F. Morales } \email{mmorales@phys.washington.edu} \affiliation{University of Washington Department of Physics \linebreak Physics Astronomy Building C525\linebreak Seattle, WA 98195}
\date{ August 9th, 2007 }



\begin{abstract}
As is common for antenna arrays in radio astronomy, the output of the MWA's correlator is the intensity measured in visibility space.  In addition, the final power spectrum will be created in visibility space.  As such, correcting for the ionosphere in visibility space instead of real space saves the computation required to inverse Fourier transform to real space and then Fourier transform back (a significant decrease in computation for systems operating in real time such as the MWA.)  In this paper, we explore this problem of correcting for ionospheric distortions in the $uv$-plane.  The mathematical formula for obtaining the unperturbed data from that reflected by the ionosphere is non-local, which in any practical application creates edge effects because of the finite nature of the $uv$-plane (section \ref{ic}).  In addition, obtaining an analytic solution for the unperturbed intensity is quite difficult, and can only be done using very specific expansions of ionospheric perturbations.  We choose one of these models (with perturbations as sinusoidal modes, section \ref{sinmodes}) and run numerical codes to further study the correction.  Numerically implementing this correction to too few orders distorts the data in such a way as to be worse than not correcting at all (section \ref{nmax}).  It is therefore critical to correct to the correct number of orders, and we present an analytic estimate for the optimal order (section \ref{theorynmax}).  This analytic estimate shows that the optimal number of orders varies with $\vec{u}$, and in particular increases as $\vec{u}$ increases along the direction of an ionospheric distorting mode.  Based on this observation, we then investigate a couple of methods which save computation (section \ref{feasible}).  These methods are (a) eliminating the intensity at values of $\vec{u}$ which require too many orders, and (b) correcting to different orders at different $\vec{u}$.  Both methods prove successful, although the first creates a loss of some precision in the real space sky.  We conclude by considering an alternate form with which to model ionospheric perturbations (section \ref{choice2}).  This alternate form was once again chosen because it lends itself to an analytic solution, but contains as many (if not more) downfalls than the original choice.
\end{abstract}

\maketitle

\section{Background on the Ionospheric Correction Operator}
\label{back}

\hspace{4 mm} We begin with the basic background on the ionospheric operator, followed by that of the ionospheric correction operator.  These sections expand upon the mathematical framework created and briefly outlined in Morales, et. al., \cite{MM}.

\subsection{Ionospheric Operator}

\hspace{4 mm} The ionospheric operator {\bf A}($\vec{\theta'}, \vec{\theta}$) is the operator which takes an unperturbed map of the sky $I(\vec{\theta})$ and maps it to a perturbed map $\tilde{I}(\vec{\theta'})$ which has been distorted by the ionosphere,
\begin{equation}
\label{Adef}
\tilde{I}(\vec{\theta'}) = {\bf A}(\vec{\theta'},\vec{\theta})I(\vec{\theta}) .
\end{equation}
Put simply, {\bf A}($\vec{\theta'},\vec{\theta}$) is a generalized coordinate change from $\vec{\theta}$ to $\vec{\theta'}$.  (The order of the arguments of {\bf A}($\vec{\theta'},\vec{\theta}$) here, and with other operators later, is indicative of the direction of change.)  

In the regime of MWA, it is a very good approximation that the mapping of angles is approximately linear with only a small deviation $\vec{\delta \theta}(\vec{\theta}, t)$,
\begin{equation}
\label{delthetadef}
\vec{\theta'} = \vec{\theta} + \vec{\delta \theta}(\vec{\theta}, t) .
\end{equation}
The perturbed intensity $\tilde{I}(\vec{\theta'})$ at $\vec{\theta'}$ is the summed contribution from the intensities $I(\vec{\theta_i})$ at all $\vec{\theta_i}$ where this relation holds; ie,
\begin{equation}
\label{discrete}
\tilde{I}(\vec{\theta'}) = \sum_i   I(\vec{\theta_i}) \delta(\vec{\theta'} - \vec{\theta_i} - \vec{\delta \theta}(\vec{\theta_i}, t)) 
\end{equation}
where $\delta$ represents the Dirac delta function.  For example, if (only) $\vec{\theta_1}$ and $\vec{\theta_2}$ are mapped to $\vec{\theta_1'}$ according to equation \ref{delthetadef}, then $\tilde{I}(\vec{\theta'_1}) = I(\vec{\theta_1}) + I(\vec{\theta_2})$.  In the limit that $\vec{\theta}$ is a continuous variable, this discrete sum for $\tilde{I}(\vec{\theta'})$ is altered to an integral, 
\begin{equation}
\label{A1}
\tilde{I}(\vec{\theta'}) = \int{ d^2\theta A(\vec{\theta'},\vec{\theta}) I(\vec{\theta}) } ,
\end{equation}
where
\begin{equation}
\label{A2}
A(\vec{\theta'},\vec{\theta}) = \delta(\vec{\theta'}-\vec{\theta}-\vec{\delta\theta}(\vec{\theta},t)) .
\end{equation}
(Notice that we've used that the magnitude of the delta-function's argument's gradient is approximately 1 here).  From this equation it is evident that in the limit of continuous $\vec{\theta}$, the ionospheric operator equation becomes
\begin{equation}
\label{Aop}
\tilde{I}(\vec{\theta'}) = {\bf A}(\vec{\theta'},\vec{\theta})I(\vec{\theta})  = \int{ d^2\theta A(\vec{\theta'},\vec{\theta}) I(\vec{\theta}) } .
\end{equation}

So far our calculations have been confined to real space, using the variables \{ $\vec{\theta'}, \vec{\theta}$ \}.  However, as is common for antenna arrays in radio astronomy, the output of the correlator for MWA will actually be the Fourier transform of the real space sky intensity, $I(\vec{u})$.  In addition, the final power spectrum of the sky will also be measured in this Fourier transfrom space (also known as \emph{visibility space}, or \emph{$uv$-space}).  Therefore, correcting for the ionosphere in real space requires inverse Fourier transforming to real space, making the correction, and then Fourier transforming back to visibility space.  The problem with this is that Fourier transforming is computationally expensive, especially for a system operating in real time, such as the MWA's \emph{Real Time System} \cite{DM}.  As such, correcting for the ionosphere in the $uv$-plane would greatly reduce computation.  We now study the nature of this $uv$-plane correction.

Let $I(\vec{u})$ represent the Fourier transform of $I(\vec{\theta})$.  Define this Fourier transform by
\begin{equation}
\label{Fourierdef}
I(\vec{u}) \equiv {\bf F}(\vec{u}, \vec{\theta})I(\vec{\theta}) \equiv \int{ d^2 \theta e^{- i \vec{u} \cdot \vec{\theta} } I(\vec{\theta}) },
\end{equation}
and the corresponding inverse Fourier transform by
\begin{equation}
\label{InverseFourierdef}
I(\vec{\theta}) \equiv {\bf F}^{-1}(\vec{\theta},\vec{u}) I(\vec{u}) \equiv \int{ \frac{d^2u}{(2\pi)^2} e^{i\vec{u} \cdot \vec{\theta}} I(\vec{u}) } .
\end{equation}
Now define {\bf A}$(\vec{u'},\vec{u})$ to be the ionospheric operator in the $uv-$plane; that is, the operator that maps the unperturbed map $I(\vec{u})$ to the perturbed map $\tilde{I}(\vec{u'})$,
\begin{equation}
\label{Audef}
\tilde{I}(\vec{u'}) = {\bf A}(\vec{u'},\vec{u}) I(\vec{u}) .
\end{equation}

One way to obtain $\tilde{I}(\vec{u'})$ from $I(\vec{u})$ is to inverse Fourier transform $I(\vec{u})$ to $I(\vec{\theta})$ using {\bf F}$^{-1}(\vec{\theta}, \vec{u})$, apply the ionospheric operator {\bf A}$(\vec{\theta'}, \vec{\theta})$ in real space to obtain $\tilde{I}(\vec{\theta'})$, and then Fourier transform $\tilde{I}(\vec{\theta'})$ to $\tilde{I}(\vec{u'})$ using {\bf F}$(\vec{u'}, \vec{\theta'})$.  In all,
\begin{equation}
\label{Audef2}
\tilde{I}(\vec{u'}) = {\bf F}(\vec{u'}, \vec{\theta'}) {\bf A}(\vec{\theta'}, \vec{\theta}) {\bf F}^{-1}(\vec{\theta}, \vec{u}) I(\vec{u}) .
\end{equation}
Comparison of this to the definition of {\bf A}$(\vec{u'},\vec{u})$ shows that
\begin{equation}
\label{Auopdef}
{\bf A}(\vec{u'},\vec{u}) = {\bf F}(\vec{u'}, \vec{\theta'}) {\bf A}(\vec{\theta'}, \vec{\theta}) {\bf F}^{-1}(\vec{\theta}, \vec{u}) .
\end{equation}
This expression may be thought of as simply a basis change of {\bf A} from \{$\vec{\theta}, \vec{\theta'}$\} to \{$\vec{u}, \vec{u'}$\}.   The three operators on the right here have all been previously given.  Plugging in these predetermined expressions (equations \ref{Aop}, \ref{Fourierdef}, and \ref{InverseFourierdef}) and simplifying as much as possible, we obtain
\begin{eqnarray}
\label{ }
\tilde{I}(\vec{u'}) & = & {\bf A}(\vec{u'},\vec{u}) I(\vec{u}) \\
& = &  \int{ \frac{d^2u}{(2\pi)^2} \left( \int{ d^2\theta e^{i(\vec{u}-\vec{u'})\cdot \vec{\theta}-i\vec{u'} \cdot \vec{\delta\theta}} } \right) I(\vec{u}) } .
\end{eqnarray}
Notice that the integral over $\vec{\theta'}$ has been evaluated by using the delta function from the expression for ${\bf A}(\vec{\theta'},\vec{\theta})$ (see equations \ref{A2} and \ref{Aop}; here again we use that the magnitude of the delta-function's argument's gradient is approximately 1).  One interesting characteristic of this expression is the non-local nature, by which finding the value of the perturbed intensity $\tilde{I}(\vec{u'})$ at one particular $\vec{u'}$ requires knowing the value of the pure intensity $I(\vec{u})$ at other $\vec{u} \neq \vec{u'}$.  This property will also appear in the ionospheric correction operator, found below.

\subsection{Ionospheric Correction Operator}
\label{ic}

\hspace{4 mm} The previous section corresponds to the operator which distorts the pure data into the perturbed data, but the reverse process is what actually interests us -- we want to correct for the effect of the ionosphere to obtain the pure data from the perturbed data.  Define {\bf A$^{T}$}$(\vec{\theta}, \vec{\theta'})$ to be this ionospheric correction operator which corrects for the influence of the ionosphere by mapping the perturbed map of the sky $\tilde{I}(\vec{\theta'})$ back to the unperturbed map $I(\vec{\theta})$,
\begin{equation}
\label{ }
I(\vec{\theta}) = {\bf A^{T}}(\vec{\theta},\vec{\theta'}) \tilde{I}(\vec{\theta'}) .
\end{equation}
The MWA will not run during periods of scintillation (at which times multiple values of $\vec{\theta}$ are perturbed to the same $\vec{\theta'}$), but will instead run during times when it is a very good approximation that the mapping from $\vec{\theta'}$ to $\vec{\theta}$ is one-to-one and approximately linear with only a small correction,
\begin{equation}
\label{ }
\vec{\theta} = \vec{\theta'} + \vec{\delta \theta'} (\vec{\theta'}, t) .
\end{equation}
The derivation of an expression for the ionospheric correction operator in the $uv$-plane ${ \bf A^{T} }(\vec{u},\vec{u'})$ follows analogously to the derivation of ${\bf A}(\vec{u'},\vec{u})$, so I'll merely quote the result:
{\small
\begin{eqnarray}
I(\vec{u}) & = & {\bf A^{T}}(\vec{u},\vec{u'}) \tilde{I}(\vec{u'}) \\
& = &  \int{ \frac{d^2u'}{(2\pi)^2} \left( \int{ d^2\theta' e^{i(\vec{u'}-\vec{u}) \cdot \vec{\theta'}-i\vec{u} \cdot \vec{\delta\theta'}} } \right) \tilde{I}(\vec{u'}) } .
\label{IUpure}
\end{eqnarray}
} 
This is the most general expression for obtaining $I(\vec{u})$ from $\tilde{I}(\vec{u'})$; proceeding further requires knowledge of the ionospheric perturbation $\vec{\delta\theta'}$.  Numerically solving for $I(\vec{u})$ using this equation is an incredibly daunting task for an arbitrary choice of $\vec{\delta\theta'}$, as it involves a double integral over all space for every value of $\vec{u}$.  Therefore, unless we find a choice of $\vec{\delta\theta'}$ which offers an analytic solution for $I(\vec{u})$, the correction for the ionosphere in the $uv$-plane will actually be more computationally expensive than the two Fourier transforms necessary to correct for the ionosphere in real space.  Unfortunately, choices of $\vec{\delta\theta'}$ which lend themselves to analytic solutions are hard to come by.  There are a couple, however, and they will be discussed in the following sections.

\section{A Specific Form for $\vec{\delta \theta'}$: Sum over Sinusoidal Modes}
\label{sinmodes}

\hspace{4 mm} The above equation for $I(\vec{u})$ contains an exponential with $\vec{\delta \theta'}$ in the exponent.  By expanding this exponential, we obtain a form for $I(\vec{u})$ which may be solved analytically for a couple of choices of $\vec{\delta \theta'}$.  To be explicit, expanding the exponential in $\vec{\delta \theta'}$,
\begin{equation}
\label{n1}
e^{-i\vec{u} \cdot \vec{\delta \theta'}} = \sum_{n=0}^{\infty}{ \frac{ (-i \vec{u} \cdot \vec{\delta \theta'})^{n}}{n!}} ,
\end{equation}
leads to
{\scriptsize
\begin{eqnarray}
\label{n2}
I(\vec{u}) & = & \int{ \frac{d^2u'}{(2\pi)^2} \left( \int{ d^2 \theta' e^{i \vec{\theta'} \cdot (\vec{u'}-\vec{u})} \sum_{n=0}^{\infty}{ \frac{ (-i \vec{u} \cdot \vec{\delta \theta'})^{n}}{n!}}    } \right) \tilde{I}(\vec{u'}) .} \nonumber \\
& = & \sum_{n=0}^{\infty} { \int{ \frac{d^2u'}{(2\pi)^2} \left( \int{ d^2 \theta' e^{i\vec{\theta'} \cdot (\vec{u'}-\vec{u})} \frac{ (-i \vec{u} \cdot \vec{\delta \theta'})^{n}}{n!}    } \right) \tilde{I}(\vec{u'}) . } }
\end{eqnarray}
} 
(Interchanging an infinite sum and an integral requires that the sum be uniformly convergent, which will be true for all choices of $\vec{\delta \theta'}$ that we choose.)  

With the above expansion, $I(\vec{u})$ may be solved analytically if we choose
\begin{equation}
\label{dtsine}
\vec{\delta \theta'} = Re( \sum_{m=1}^{M} ia_m \vec{b}_m e^{-i \vec{b}_m \cdot \vec{\theta'} }  ).
\end{equation}
Here, the $\vec{b}_m$ are chosen to be purely real, but the $a_m$ are allowed to assume complex values.  Physically, this choice of $\vec{\delta \theta'}$ corresponds to modeling the integral along the line of sight of the ionosphere's electron density $n_e(\vec{\theta'},h)$ ($h$ is the distance along the line of sight) as a sum over sinusoidal modes, 
\begin{equation}
\label{density}
\int n_e(\vec{\theta'}, h) dh  \propto \sum_{m=1}^{M} e^{-i \vec{b}_m \cdot \vec{\theta'} }.
\end{equation}
The reflection by the ionosphere is then related to this by
\begin{equation}
\label{densityreflect}
\vec{\delta \theta'} \propto \nabla_{\vec{\theta'}} \left(\int n_e(\vec{\theta'}, h) dh \right) \propto \sum_{m=1}^{M} -i \vec{b}_m e^{-i \vec{b}_m \cdot \vec{\theta'} }, 
\end{equation}  
where $\nabla_{\vec{\theta'}}$ represents the two-dimensional gradient with respect to $(\theta'_x, \theta'_y)$.  (The actual shift, of course, is a Hermitian observable, so only the real part of this is included in $\vec{\delta \theta'}$.)  Notice that this choice ultimately stems from our decision to model density fluctuations from the ionosphere as sinusoidal modes.  Actually, any orthonormal basis would have sufficed here.  Once again, this particular choice was made because it allows an analytic solution for the unperturbed intensity $I(\vec{u})$.  (An alternate choice which likewise offers an analytic solution will be briefly discussed later on in section \ref{choice2}.)

With this choice of $\vec{\delta \theta'}$, the intensity $I(\vec{u})$ becomes
{\footnotesize
\begin{eqnarray}
I(\vec{u}) = & & \sum_{n=0}^{\infty} \int \frac{d^2u'}{(2\pi)^2} \int \frac{ d^2 \theta' }{n!} \left[ e^{i\vec{\theta'} \cdot (\vec{u'}-\vec{u})}  \right. \mbox{ \hspace{5 mm} } \\
& & \mbox{ {\textbf X} } \left. \left( -i \vec{u} \cdot Re( \sum_{m=1}^{M} ia_m \vec{b}_m e^{-i \vec{b}_m \cdot \vec{\theta'} }  )    \right)^{n} \right]  \tilde{I}(\vec{u'}) .
\end{eqnarray}
} 
The math leading to a solution for $I(\vec{u})$ may be found in the appendix, section \ref{math1}.  Very briefly, the integral is solved by conveniently redefining the ionospheric modes $\{ a_m, \vec{b}_m \} $ (as given below), performing a multinomial expansion on the term raised to the power $n$, recognizing that the final product of this expansion leaves the $\vec{\theta'}$ integral in the form a delta function, and then using that delta function to evaluate the integral over $\vec{u'}$.  The final solution is
\begin{equation}
\label{finalIU}
I(\vec{u}) = \sum_{n=0}^{\infty} \sum_{l_1, l_2, ... , l_{2M}}' \left( \prod_{q=1}^{2M} \frac{ ( c_q \vec{u} \cdot \vec{d}_q)^{l_q}}{l_q!} \right) \tilde{I}(\vec{u} + \sum_q l_q \vec{d}_q ) ,
\end{equation}
where
\begin{equation}
\label{c_qdef}
 c_q = \left\{ \begin{array}{ll}
         \frac{1}{2} a_q & \mbox{if $q <  M+1$}\\
         \frac{1}{2} a_{q-M}^* & \mbox{if $q \geq M+1 $}\end{array}, \right. 
\end{equation} 
\begin{equation}
\label{d_qdef}
 \vec{d}_q = \left\{ \begin{array}{ll}
         \vec{b}_q & \mbox{if $q < M+1$}\\
         - \vec{b}_{q-M} & \mbox{if $q \geq M+1$},\end{array} \right. 
\end{equation} 
and the summation over all $l_q$ (with $2M \geq q \geq 1$) is a restricted sum such that $\sum_q l_q = n$ and $l_q \geq 0$.  This equation might look a little daunting, but it may be thought of simply as the addition of many delta functions of varying amplitudes, with those further from the point $\vec{u}$ in question tending to contribute less to the sum.  (This is, in fact, similar to what one sees with intermodulation distortion).  Notice that although there are $M$ modes distorting the sky, the sums and products above involve $2M \geq q \geq 1$.  Thus, there appear to be $2M$ \emph{effective modes} distorting the sky.  This factor of 2 comes from the constraint that $\vec{\delta \theta'}$ be real, as may be more easily seen by following the math provided in the appendix, section \ref{math1}.

Another important feature of this solution is that it is inherently non-local, with the corrected intensity $I(\vec{u})$ at a given $\vec{u}$ depending on the values of the perturbed intensity $\tilde{I}(\vec{u'})$ at the appropriate neighboring points $\vec{u}' = \vec{u} + \sum_q l_q \vec{d}_q $.  This non-locality, which is also evident in the most general form for $I(\vec{u})$ (equation \ref{IUpure}), will create edge effects because the $uv$-plane is finite in all practical applications, as will be more easily seen and understood later in section \ref{trunc10}.

It should also be pointed out that this equation for $I(\vec{u})$ is the result of a double expansion:  the Taylor-series expansion of the exponential containing $\vec{\delta\theta'}$ (see equation \ref{n1}), which is now evident in the summation over $n$, and the expansion of the ionospheric perturbation $\vec{\delta\theta'}$ itself into sinusoidal modes (see equation \ref{dtsine}), which is now evident in the restricted sum over $l_q$.  Throughout this paper we will assume that this second expansion is ``perfect"; that is, we will assume that we are able to perfectly model the ionosphere with the $M$ modes that we assume are provided for us.  We will instead study the errors created by truncating the first expansion.

\section{Potential Problem:  Computational Feasibility of the Two Expansions}
\label{numterms}

\hspace{4 mm} As previously stated, the main goal of the $uv$-plane correction is to correct for the ionosphere in a less computationally intensive manner than that required for the real space correction.  Our final expression for $I(\vec{u})$, however, contains an infinite sum over $n$.  Clearly, making the $uv$-plane correction computationally feasible will require truncating this sum after a finite number of terms.  The next section will explore the effect of such a truncation.

But even truncating this sum over $n$ cannot guarantee the computational feasibility of the $uv$-plane correction because of the second expansion over sinusodial modes and its resulting restricted summation.  More specifically, the number of terms in the restricted sum over all possible combinations of $l_q$ such that $\sum_q l_q = n$ may be calculated through the following trick:  if $2M$ represents the total number of effective modes, then consider the problem of arranging $n$ balls and $2M-1$ partitions in a straight line.  Here the $i$-th partition marks the stopping point where $l_i$ ends and $l_{i+1}$ begins.  For example, if for a particular arrangement 7 balls lie between the 4th and 5th partitions, then $l_5$ = 7 for that arrangement.  The total number of ways to arrange these $n+2M-1$ objects is $(n+2M-1)!$.  Of course, exchanging the position of any two of the same object (ball or partition) does not lead to a different arrangement, so the total number of terms in the restricted sum such that $\sum_{q=1}^{2M} l_q = n$ is given by
{\small
\begin{equation}
\left( \mbox{\# of terms in the sum \hspace{2 mm} } \sum_{q=1}^{2M} l_q = n \right) = \frac{(n + 2M - 1)!}{n!(2M-1)!} .
\end{equation}
} 

As an example, suppose that we wish to calculate this sum for 10 modes ($2M=20$ effective modes) to the 40th order in $n$.  Using the above formula, we calculate that such a sum has approximately 70 trillion terms.  From this we see that the $uv$-plane correction is only computationally feasible if the number of modes necessary to model the sky $M$ \emph{and} the number of orders necessary in the expansion of the exponential $n$ are relatively small. 

\section{Computational Results:  Truncating the Infinite Sum Over $n$}
\label{nmax}
\hspace{4 mm} As stated in the previous section, the $uv$-plane correction is only computationally feasible if we truncate the infinite sum over $n$.  At this point we pause to study the results of truncating this sum after a finite order of correction, $n=n_{max}$.

\subsection{Simple Sky Model}

\begin{figure}
\begin{center}
\PSbox{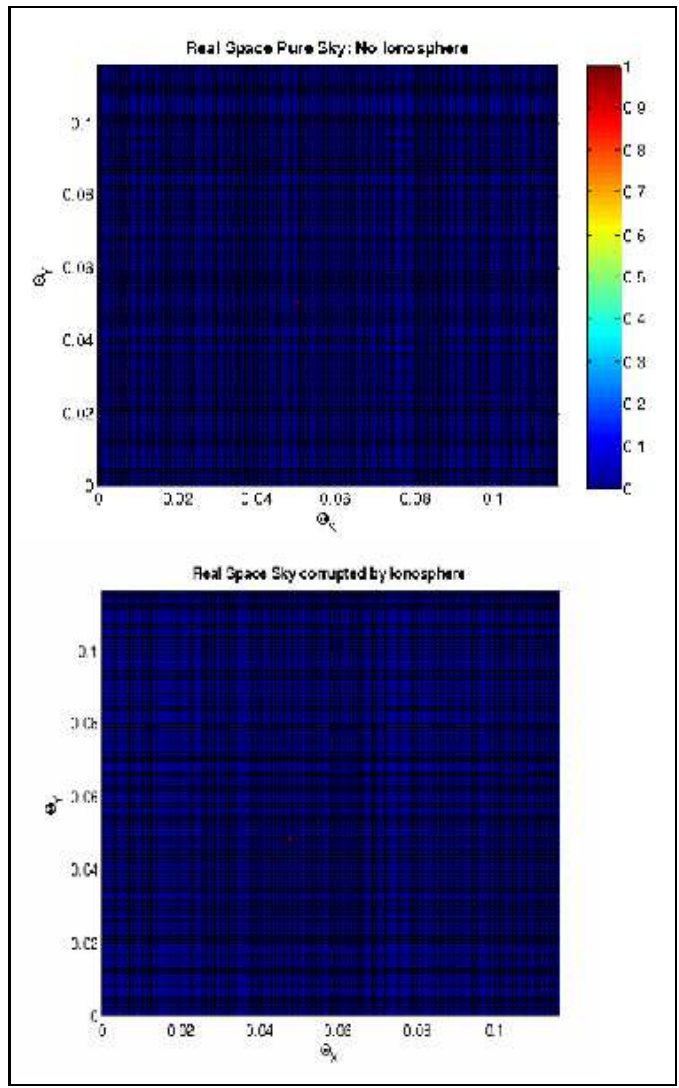}{3.4in}{5.56in}
\caption{\emph{Top Panel}  A simple model of a sky (referred to as \emph{simple sky} throughout this paper) consisting of a single source; ie, the value at each pixel is set to 0.0 except at one pixel, where it is set to 1.0.  Notice that this simple model has no side lobes.  \emph{Bottom Panel}  The simple sky from the top panel, but shifted by a single ionospheric mode, as given by equation \ref{dtsine} with $a_m = 1.0 \times 10^{-5}$ radians squared and $\vec{b_m}$ with a magnitude of 378.0 inverse radians and oriented in the $\pi/4$ direction with respect to the $\theta_x$ axis.  It is difficult to see from these plots, but the single source has shifted in the $\pi/4$ direction by a few pixels and still has an intensity of 1.0.  The axes of both graphs are measured in radians.} 
\label{pureandbad}
\end{center}
\end{figure}

\begin{figure}
\begin{center}
\PSbox{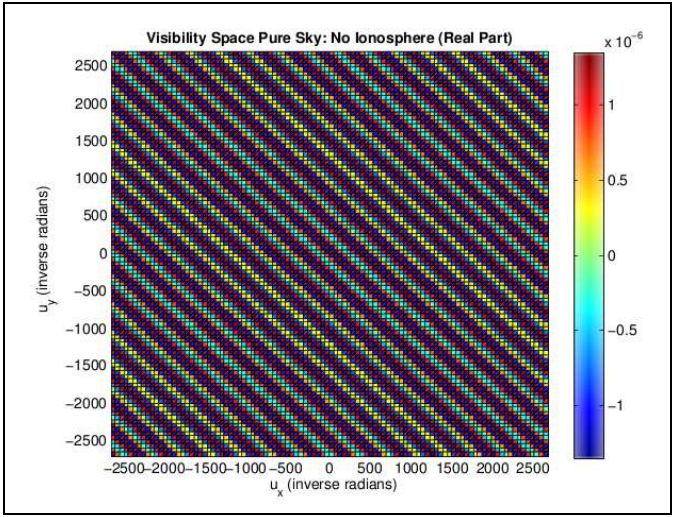}{3.4in}{2.59in}
\caption{This is the pure visibility sky; ie, the Fourier transform of the simple unperturbed sky shown in the top panel in figure \ref{pureandbad}.  The dimensions of the axes are inverse radians, and the third dimension represented by the color scale is the real part of the intensity.  Notice from these two panels that the pure, uncorrupted intensity in visibility space is a simple sinusoid, which is what one would expect for the Fourier transform of a delta-function.  The important point to take away from this plot is that the magnitude intensity is on the order of $10^{-6}$ for our simple sky, and is the same for each $uv$-pixel (although only the real part is plotted here).  All other $uv$-plots in this paper will have the magnitude of the intensity (and not just the real part) for the third, colored dimension. }
\label{uv}
\end{center}
\end{figure}

\hspace{4 mm}  In order to study the qualitative effects of truncating the sum over $n$, we created a simple sky and perturbed it with a simple mode, and then used numerical code to correct for this (known) perturbation in the $uv$-plane by using the mathematical formula found above in equation \ref{finalIU}.

The initial, unperturbed sky is shown in the top panel of figure \ref{pureandbad}.  We will refer to this sky throughout this paper as the \emph{simple sky}.  It contains a $101 \times 101$ pixel array with a spacing of 4 arcmin between pixels, which is the approximate resolution we expect for the final MWA array (The axes in this and all the following real space figures are labeled in radians).  This pure sky is a single source sky:  The value of the intensity at all pixels is set to 0.0 except at one pixel where the value 1.0.  (The important qualitative results found below would not be altered by including side lobes, so we will leave them out to keep things simpler.)

Figure \ref{uv} shows the real part of the intensity in the $uv$-plane for this pure sky plotted in the third direction (which is determined by the color scale) and demonstrates that the uncorrupted, $uv$-sky is a simple sinusoid (as one would expect for the Fourier-transform of a delta function).  In contrast to this plot, the color scale for all the following $uv$-space plots is representative of the magnitude of the intensity $I(\vec{u})$ in the $uv$-plane (although $I(\vec{u})$ is complex, the important $uv$-plane results found below do not require phase information to understand graphically).  The important feature to take away from this plot is that the absolute magnitude of the intensity is constant and of the order $10^{-6}$ at all points in the $uv$-plane (although we've only plotted the real part here).

We then perturb this simple sky with a rather strong mode, as shown in the bottom panel of figure \ref{pureandbad}.  This distorting mode has $|a_m| = 1.0 \times 10^{-5}$ radians squared and $\vec{b}_m$ with a magnitude of 378.0 inverse radians and oriented in the $\pi/4$ direction as measured from the $\theta_x$ axis.  (These values were chosen for the sole reason that they produce a strong shift of a few pixels (tens of arcminutes), and thus accentuate the qualitative features of the ionospheric correction as seen below.  A more realistic distortion will be discussed later in section \ref{realistic}.)   Notice that the intensity is still 1.0 at exactly one pixel and 0.0 at every other pixel, but now the location of this pixel has slightly shifted in the $\pi/4$ direction.  (It should be noted that this ionospheric shift was applied to the continuous sky with a delta function at one point, and not to its pixelized represention shown in the top panel of figure \ref{pureandbad}).  Although we have not included the plot, the magnitude of the intensity in the perturbed $uv$-plane remains of the order of $10^{-6}$, its value in the pure visibility sky.

\subsection{Truncation Through 10 Orders for our Simple Sky}
\label{trunc10}

\begin{figure}
\begin{center}
\PSbox{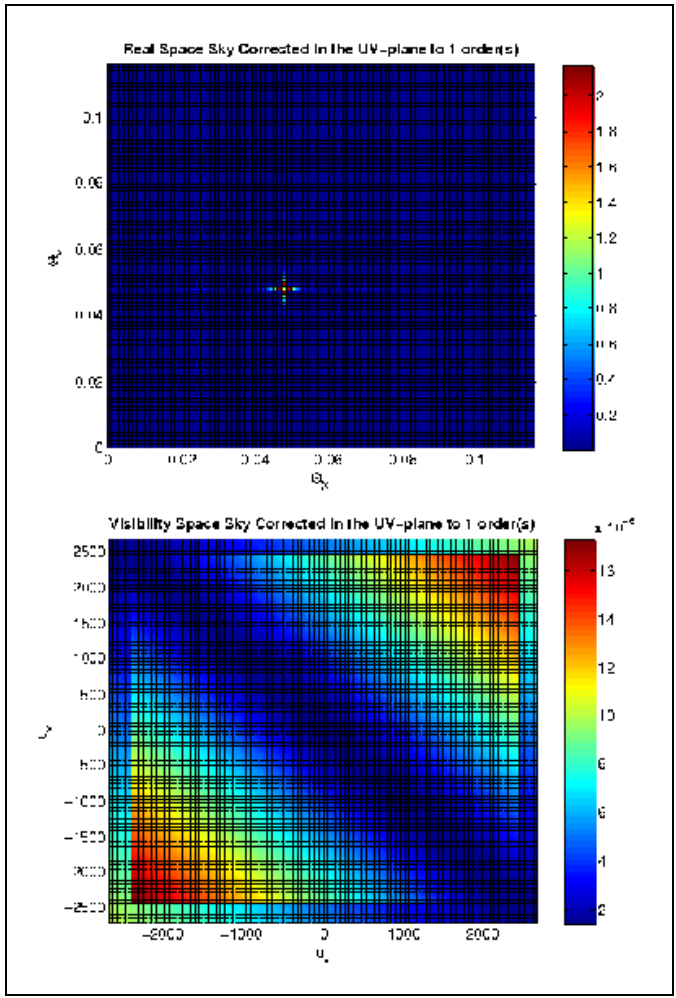}{3.4in}{5.1in}
\caption{The results from correcting to first order in the $uv$-plane.  \emph{Top Panel} The real space sky (the axes have dimensions of radians): After correction to first order, the star remains at its perturbed location.  The most important feature of this graph is the intensity of the sky.  In both the pure and corrupted skies of figure \ref{pureandbad} this intensity had been 1.0 at one pixel and 0.0 at all others.  Here, it takes a maximum value of approximately twice that.  \emph{Bottom Panel}  The visibility space sky (the axes have dimensions of inverse radians): This plot has two important features: 1) The maximum in the intensity, which has now increased by an order of magnitude near the extreme lower left and upper right corners compared to its value in the pure visibility sky (figure \ref{uv}) and corrupted visibility sky.  2) The border around the edge, which is caused by the non-local nature of the $uv$-plane correction (see equation \ref{finalIU}) combined with the finite nature of our numerical $uv$-plane.  More specifically, points in this border strip are not entirely corrected because part of their correction requires points in the $uv$-plane outside of the numerical grid.  }
\label{order1}
\end{center}
\end{figure}

\hspace{4 mm} We now begin correcting for this simple one-mode distortion using various values of $n_{max}$.  We (quite naturally) begin with the first order correction, $n_{max} = 1$ ($n_{max} = 0$ leads to no correction, see equation \ref{finalIU}).  After correcting to first order in the $uv$-plane, we inverse Fourier transform back to the real space sky shown in the top panel of figure \ref{order1} in order to determine the effect that this first order correction has had on the real sky (in particular, we would like to know whether it has successfully shifted the single source back to its unperturbed location).  As it turns out, the correction to one order has not shifted the star from its perturbed location.  The cross-like pattern of the star is somewhat interesting, but what is most important about this figure is that the maximum in the intensity has now doubled from 1.0 to 2.0.

A potential clue to this behavior is found by studying the visibility space sky corrected to first order, as shown in the bottom panel of figure \ref{order1}.  From this figure we see that the first order correction in the $uv$-plane has created an increase of an order of magnitude in the absolute value of the intensity at those points in the $uv$-plane furthest from the origin.

Another important feature of this figure (although, as it turns out, it is not the cause of the increase in the real space intensity) is the ring around the outside edge of the figure.  This ring is caused by the previously mentioned fact that the correction in the $uv$-plane is non-local (see equation \ref{finalIU}), combined with the finite nature of our numerical $uv$-plane.  More specifically, points near the edge of our $uv$-plane may not obtain the full correction in visibility space, because doing so requires pulling values of the intensity that are off the edge of the grid.  Therefore, values near the edge are never fully corrected.  We will later see that the result of this is a small spreading of the initial source (ie, a loss of precision), in real space.

\begin{figure}
\begin{center}
\PSbox{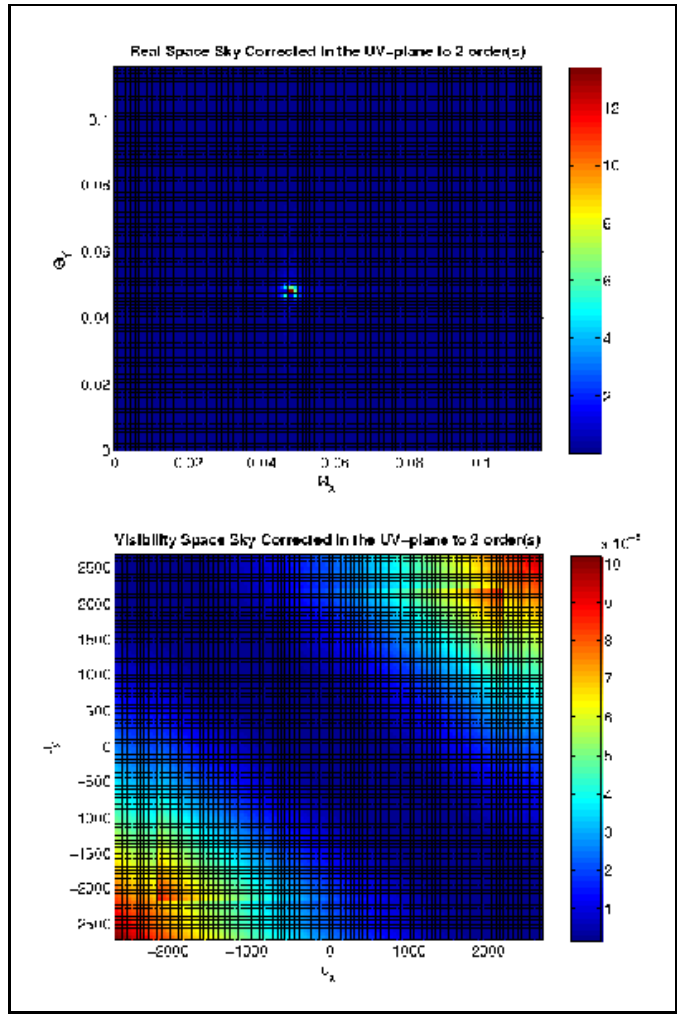}{3.4in}{5.19in}
\caption{The results from correcting to second order in the $uv$-plane.  \emph{Top Panel} The real space sky (the axes have dimensions of radians):  Notice that the maximum in the intensity is now approximately 12 times that of the pure and corrupted skies of figure \ref{pureandbad}.  (Recall that the real space intensity after the first order correction was only 2 times too large).  \emph{Bottom Panel}  The visibility space sky (the axes have dimensions of inverse radians):  Notice that the maximum in the visibility sky intensity is 10 times the maximum for the first order correction (figure \ref{order1}, bottom panel) and 100 times the maximum for the pure simple visibility space sky (figure \ref{uv}). }
\label{order2}
\end{center}
\end{figure}

\begin{figure}
\begin{center}
\PSbox{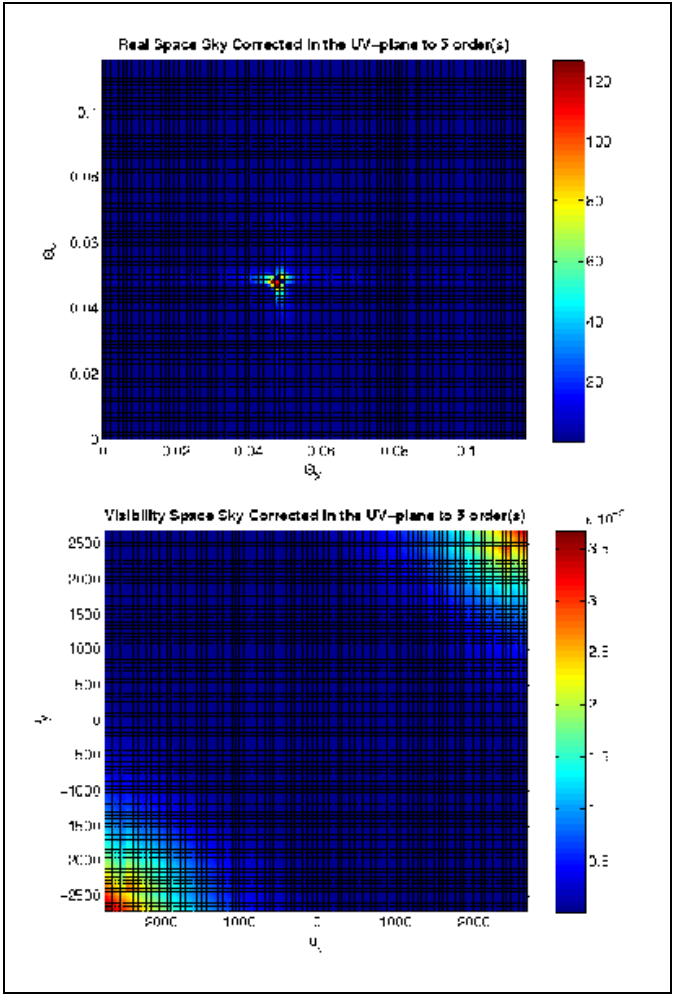}{3.4in}{5.1in}
\caption{The results from correcting to fifth order in the $uv$-plane.  \emph{Top Panel} The real space sky (the axes have dimensions of radians):  The maximum in the intensity continues to increase compared with the pure and corrupt skies (figure \ref{pureandbad}) and lower order corrections (figures \ref{order1} and \ref{order2}, both top panels), and is now approximately 120 times too large.   \emph{Bottom Panel}  The visibility space sky (the axes have dimensions of inverse radians):  The maximum in the intensity continues to increase near the lower left and upper right extremes, and is 1000 times larger than the maximum in the pure visibility sky (figure \ref{uv}).  }
\label{order5}
\end{center}
\end{figure}

\begin{figure}
\begin{center}
\PSbox{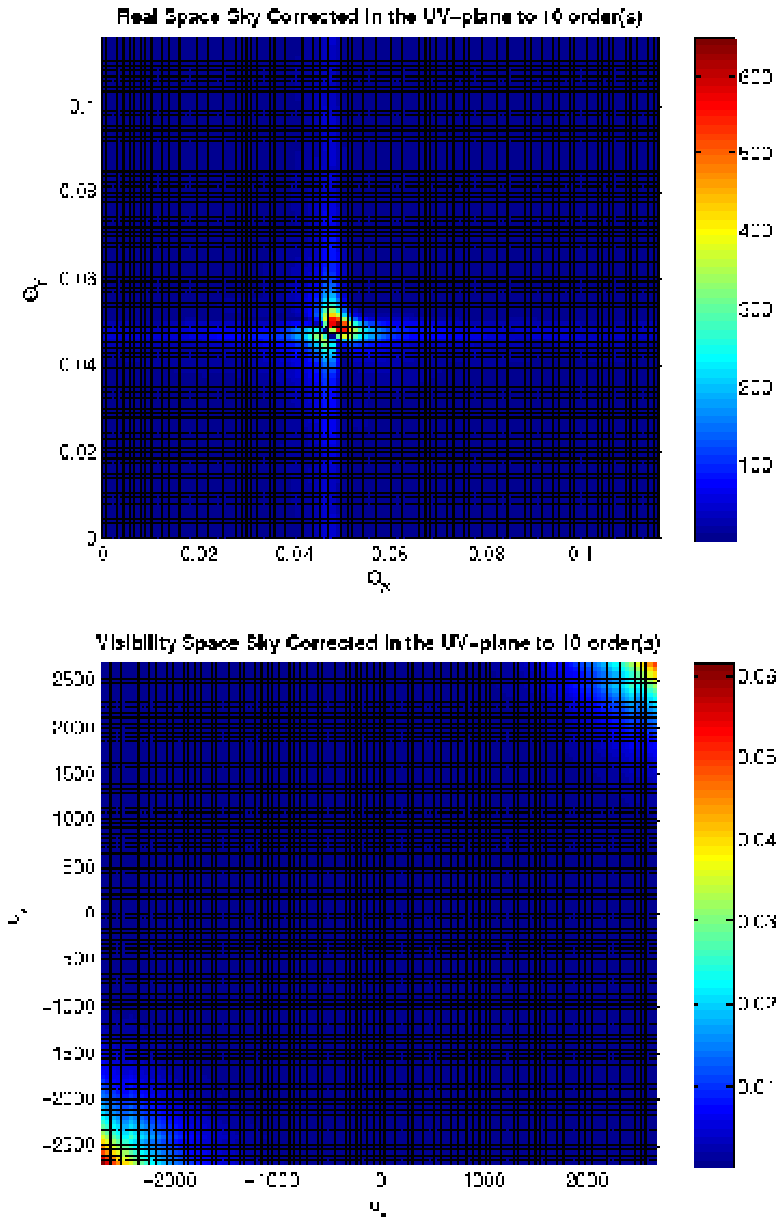}{3.4in}{5.06in}
\caption{The results from correcting to tenth order in the $uv$-plane.  \emph{Top Panel} The real space sky (the axes have dimensions of radians):  The upward trend in the maximum of the intensity increases, and the maximum in the intensity is now approximately 600 times its value in the pure and uncorrected corrupt skies (figure \ref{pureandbad}).  In addition, the star, initially a single-pixel point source, is now spread over an appreciable number of pixels.  \emph{Bottom Panel}  The visibility space sky (the axes have dimensions of inverse radians):  Likewise, the maximum in the visibility space intensity continues to increase in two of the extreme corners, and is now approximately 10,000 times what is was in the pure visibility space sky (figure \ref{uv}) and uncorrected corrupt sky. }
\label{order10}
\end{center}
\end{figure}

We now continue on to the second order correction ($n_{max} = 2$).  After inverse Fourier transforming, we obtain the real space sky shown in figure \ref{order2} (top panel).  From this figure we see that the maximum in the intensity has increased even more, and is now 12 times its unperturbed  and uncorrected value.  The visibility sky after two orders of correction, as shown in the bottom panel of figure \ref{order2}, has a maximum in the intensity that is now 100 times the value of the pure $uv$-sky.

Correcting to higher orders, we see that the problem with intensities that are too high not only persists, but continues to get worse.  After 5 orders, the intensity in the real space sky (figure \ref{order5}, top panel) is 120 times too large, and that in the visibility space sky (figure \ref{order5}, bottom panel) is $10^{3}$ times too large; after 10 orders, the intensity in the real space sky (figure \ref{order10}, top panel) is 600 times too large, and that in the visibility space sky (figure \ref{order10}, bottom panel) is $10^{4}$ times too large.  In addition to the increase in the maximum in real space intensity, the source is also beginning to spread out and look less like a single point source.

\subsection{Making Sense of the Bizarre Behavior of the $uv$-Plane Correction}

\hspace{4 mm}  Before continuing, let's pause to develop an intuition of what is happening here.  Consider a simple exponential, 
\begin{equation}
\label{simpleexp}
x = e^{7123i} \approx -.53232 - .84654i .
\end{equation}
Suppose that we want to approximate this exponential using a Taylor expansion, 
\begin{equation}
\label{te}
x = \sum_{n=0}^{\infty} \frac{(7123i)^n}{n!} .
\end{equation}
It makes sense that a decent approximation to the original exponential should be possible by truncating this sum after a finite number of terms.  But how many terms are necessary?

Let's first consider the zeroth order approximation, in which only the $n = 0$ term is kept,
\begin{equation}
\label{x0}
x_0 = 1.
\end{equation}
Notice that the zeroth order approximation gives us the right magnitude of 1, but all information about the phase has been lost.  If we instead correct to first order, we obtain
\begin{equation}
\label{x1}
x_1 = 1 + 7123i .
\end{equation}
This is not even close to the correct answer -- not only does this not contain the correct phase, but the magnitude is now not even close to being correct.  Correcting to second order gets us even further from the correct answer,
\begin{equation}
\label{x2}
x_2 = -25368563.5 + 7123i.
\end{equation}

This pattern continues for higher orders as well.  In fact, the approximation won't begin to look decent until $n \approx 7123$.  Even more relevant to our observations in the previous section, notice that adding subsequent terms to the approximation does not necessarily make the approximation better until $n \approx 7123$.  Before then, adding subsequent terms actually makes the approximation worse.  
Drawing from these observations, we expect that the trend we've seen so far is the result of under correcting in the $uv$-plane, and that by going to more and more orders we will eventually obtain a decent correction.

\subsection{Higher Order Corrections, 10+ orders}

\begin{figure}
\begin{center}
\PSbox{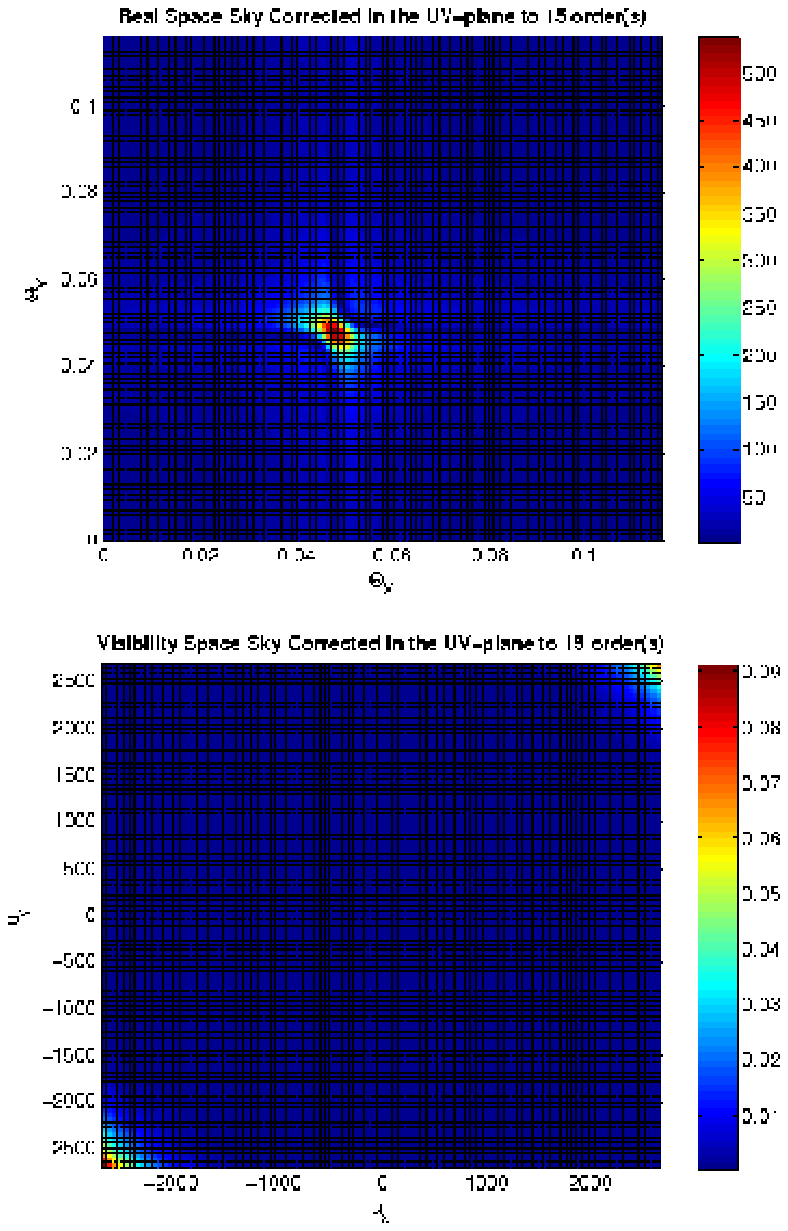}{3.4in}{5.19in}
\caption{The results from correcting to fifteenth order in the $uv$-plane.  \emph{Top Panel} The real space sky (the axes have dimensions of radians):  The maximum in the intensity has decreased relative to its 10th order counterpart (figure \ref{order10}, top panel), from approximately 600 to approximately 500 times too large relative to the pure and uncorrected corrupt skies (figure \ref{pureandbad}), but the source continues to spread in extent.  \emph{Bottom Panel}  The visibility space sky (the axes have dimensions of inverse radians):  While the real space sky showed a decrease in intensity compared to the 10th order correction, the maximum in the visibility sky intensity is still approximately 10,000 times too large compared to its pure counterpart (figure \ref{uv}).  While the visibility sky intensity does not look much better in this sense, it at least does not appear to be getting worse.  }
\label{order15}
\end{center}
\end{figure}

\begin{figure}
\begin{center}
\PSbox{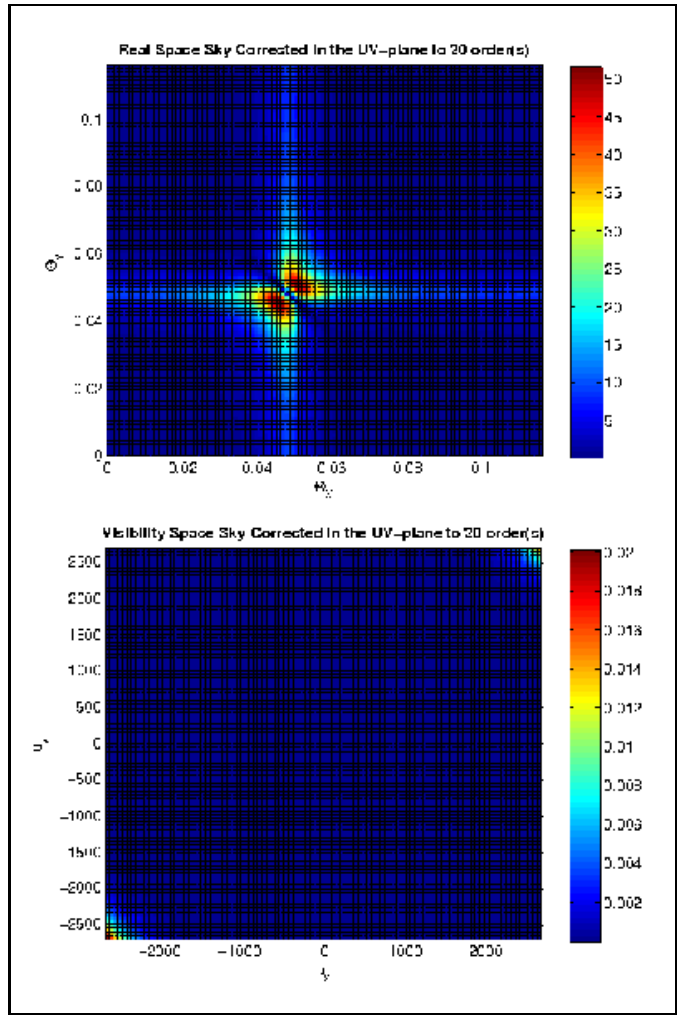}{3.4in}{5.20in}
\caption{The results from correcting to twentieth order in the $uv$-plane.  \emph{Top Panel} The real space sky (the axes have dimensions of radians):  The maximum in the intensity continues to decrease now as we continue to higher orders of correction.  The maximum is now only approximately 50 times its value for the pure and uncorrected corrupt skies (figure \ref{pureandbad}), down by a factor of 10 compared to the fifteenth order correction (figure \ref{order15}, top panel).  But while the maximum intensity looks better, the star has spread even further in extent.  \emph{Bottom Panel}  The visibility space sky (the axes have dimensions of inverse radians):  The maximum in the visibility space sky is comparable to the tenth (figure \ref{order10}, bottom panel) and fifteenth (figure \ref{order15}, bottom panel) order values at approximately 10,000 times larger than the pure (figure \ref{uv}) and uncorrected corrupt visibility space skies.   }
\label{order20}
\end{center}
\end{figure}

\begin{figure}
\begin{center}
\PSbox{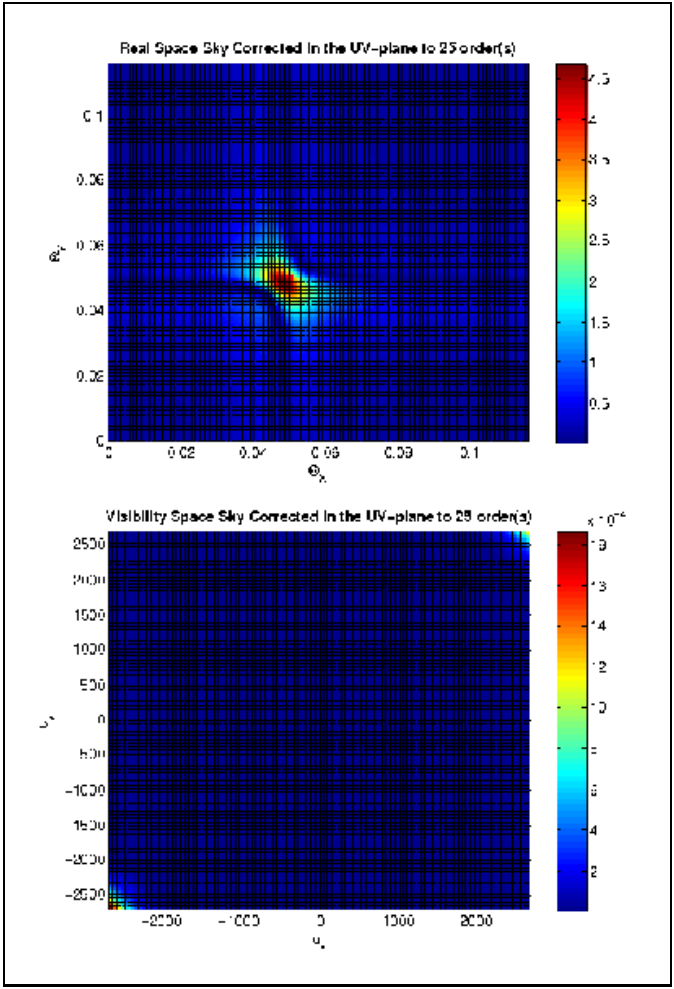}{3.4in}{5.06in}
\caption{The results from correcting to twenty-fifth order in the $uv$-plane.  \emph{Top Panel} The real space sky (the axes have dimensions of radians):  The maximum in the intensity of the real space sky continues to decline, and is now only 4 or 5 times larger than the actual pure simple sky value (figure \ref{pureandbad}).  The star, however, still does not resemble a point source.   \emph{Bottom Panel}  The visibility space sky (the axes have dimensions of inverse radians):  The maximum in the intensity of the visibility space sky is now beginning to decline, and is now only 1000 times its value in the pure visibility space sky (figure \ref{uv}), compared to 10,000 times too large for the twentieth order correction (figure \ref{order20}).     }
\label{order25}
\end{center}
\end{figure}

\begin{figure}
\begin{center}
\PSbox{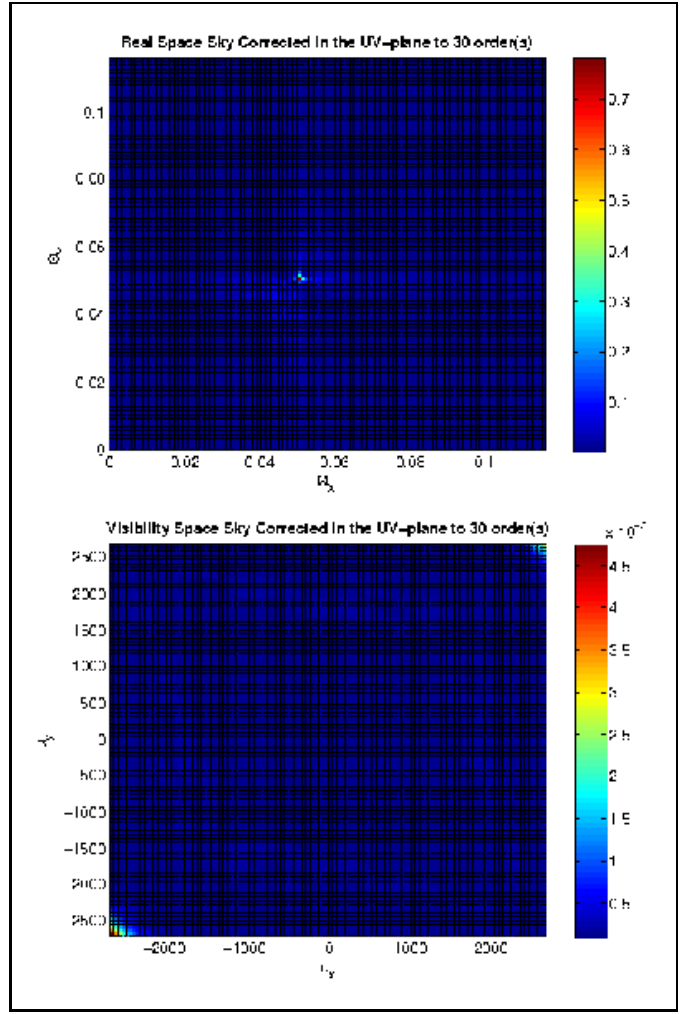}{3.4in}{5.19in}
\caption{The results from correcting to thirtieth order in the $uv$-plane.  \emph{Top Panel} The real space sky (the axes have dimensions of radians):  The single source from our simple sky (figure \ref{pureandbad}) now once again looks like a simple source, but with its intensity spread over a few neighboring pixels.  The maximum in the intensity is now of the same order of magnitude as that of the pure sky (figure \ref{pureandbad}), and is located at the same pixel.  \emph{Bottom Panel}  The visibility space sky (the axes have dimensions of inverse radians):  The maximum in the visibility space intensity, which is located at the lower left and upper right corners as always, is now only an order of magnitude higher than that of the pure visibility sky (figure \ref{uv}).  }
\label{order30}
\end{center}
\end{figure}

\begin{figure}
\begin{center}
\PSbox{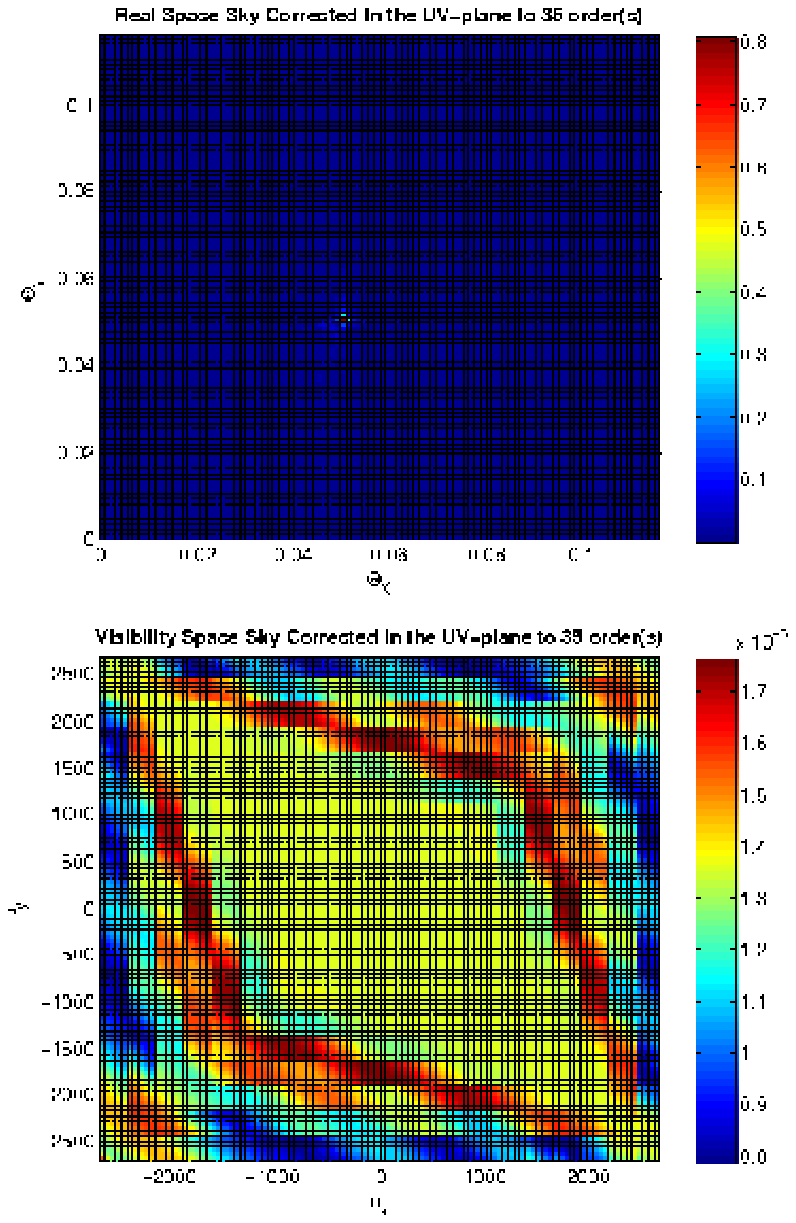}{3.4in}{5.1in}
\caption{ The results from correcting to thirty-fifth order in the $uv$-plane.  \emph{Top Panel} The real space sky (the axes have dimensions of radians):  This real space sky shows minor improvements over the real space sky corrected to 30 orders in the $uv$-plane (figure \ref{order30}).  For both, the source has been successfully shifted back to its correct starting pixel, as shown in the top panel of figure \ref{pureandbad}, at the cost of minor spreading to the neighboring pixels.  \emph{Bottom Panel}  The visibility space sky (the axes have dimensions of inverse radians):  The entire visibility space sky (including the extremes) now has an intensity of the order $10^{-6}$, which is the order of the intensity for the pure visibility space sky (figure \ref{uv}).  Notice that since the most extreme pixels are no longer much more intense than the others, we now see the pattern caused by the non-locality of the correction and finite nature of the grid (see section \ref{trunc10}).  }
\label{order35}
\end{center}
\end{figure}

\hspace{4 mm}  We now verify that this intuition is correct by studying higher order corrections.  If correct, we expect to see the results gradually get better.  We now consider the fifteenth order correction, $n_{max}$ = 15, as shown in figure \ref{order15}.  The real space sky after 15 orders (top panel) is now only 500 times too intense (versus 600 for 10 orders), while the visibility space sky (bottom panel) is still approximately $10^{4}$ times too intense.  From this it is unclear that things are getting better, but in the very least the intensities are not getting worse.  The shape of the source, however, continues to grow further from a point source.

Moving on to $n_{max} = 20$ (figure \ref{order20}) is a bit more reassuring.  The maximum in the real space sky intensity is now approximately only 50 times its actual value (top panel), although the maximum in the visibility space intensity is still four orders too high (bottom panel).  The gradual improvement continues when we skip ahead to 25 orders (figure \ref{order25}).  The real space intensity is now only 4 or 5 times too large (top panel), while the visibility space intensity has now dropped to 1000 times too large (bottom panel).  The shape of the source, however, continues to grow worse.    

Skipping ahead next to 30 orders of correction shows a dramatic improvement.  The top panel of figure \ref{order30} shows that the single source now appears to be a single source of the right order of magnitude in intensity.  And in addition, the most intense pixel is now located exactly where it was for the pure sky, so the $uv$-plane correction has (at least in terms of location of the max) successfully corrected for the shift by the ionosphere.  The bottom panel of figure \ref{order30} shows that the maximum in the visibility space sky intensity (which, as always, occurs near the edge of the grid) is now only an order of magnitude too big.  It appears as if we've gone over the hump, and are now on our way to decent results.

The correction to 35 orders shows minor improvement in real space (figure \ref{order35}, top panel).  In the visibility space sky (figure \ref{order35}, bottom panel), however, the entire grid now has the correct order of magnitude of $10^{-6}$, including the most extreme pixels.  We therefore now see some of the finer patterns caused by the non-locality of the correction and finite nature of the $uv$ grid (as mentioned previously in section \ref{trunc10}), which had previously been hidden by the extreme intensities at the corners.

It should be noted that the most intense pixel in this fully corrected real space sky in the top panel of figure \ref{order35} still lies at the location of the single source in the original, pure sky.  In other words, the $uv$-plane correction has successfully shifted the reflected source back to its initial position, at the cost of minor spreading over a few neighboring pixels.  This spreading, which cannot be eliminated by correcting to still higher orders, is caused by the finite nature of the $uv$-plane and thus cannot be avoided.

It turns out that corrections to higher orders show negligible improvement over the correction to 35 orders, so the resulting skies, identical to those of figure \ref{order35}, are not shown. 

\begin{figure}
\begin{center}
\PSbox{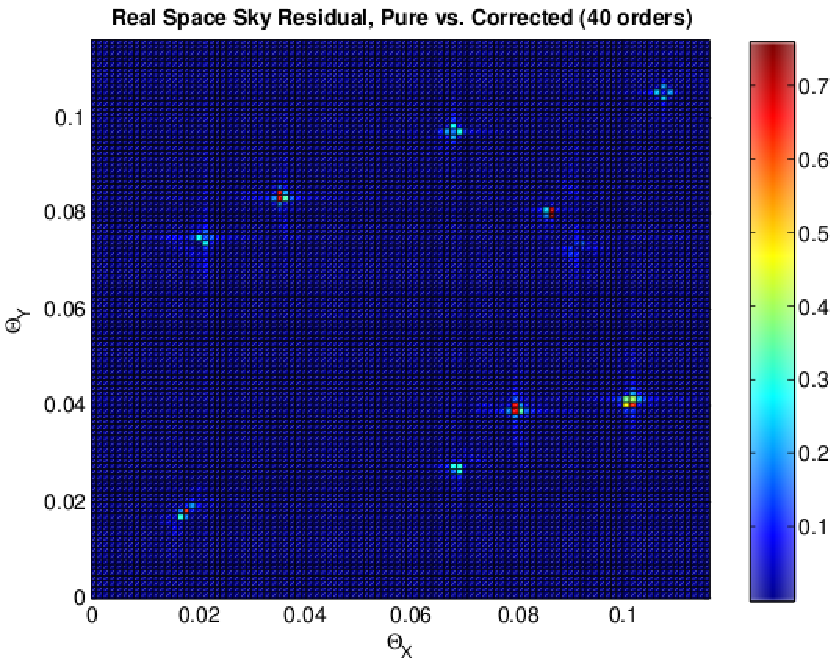}{3.4in}{2.69in}
\caption{\emph{10 Star Sky}  This plot shows the residual between a pure real space sky with 10 stars, and the real space sky obtained after perturbing this pure sky and then correcting in the $uv$-plane to 40 orders.  The ionospheric perturbation was the same used for all the other figures in this section, figures \ref{pureandbad} to \ref{order35}.  This figure shows the errors typical with the $uv$-plane correction.  The top right cross represents a star shifted back to the correct position, but with an intensity spread over a few pixels.  Stars with two consecutive high intensity pixels represent stars shifted back to pixels adjacent to their correct starting pixels.  (Typically, stars are shifted 3 or 4 pixels by the ionosphere, so the $uv$-plane correction has provided some improvement with these stars as well).       }
\label{10stars}
\end{center}
\end{figure}

So far we have only used a particularly simple sky with one star.  Figure \ref{10stars} shows (the absolute value of) the residual between a more complicated pure sky with 10 stars, and the real space sky obtained after perturbing this pure sky and then correcting in the $uv$-plane to 40 orders.  For comparison's sake the perturbation used here was the same ionospheric mode used above to perturb the simple sky of figure \ref{pureandbad} (ie, that used throughout this section).  This figure shows the kinds of errors we may expect from the $uv$-plane correction.  The residual from the star on the top right shows a light cross pattern, indicative of a spreading of the source caused by the process of perturbing the star and then applying the $uv$-plane correction.  Places in the plot with two consecutive pixels of high intensity represent stars which were not shifted back to exactly the same pixel that they started at, but rather to a neighboring pixel.  Recall that the stars are typically initially shifted 3 or 4 pixels by the ionosphere, so the $uv$-plane correction is still providing some improvement with these stars.

\section{Analytic Estimate for $n_{max}$}
\label{theorynmax}

\hspace{4 mm} There are two main points to be taken from our analysis so far.  First, from section \ref{numterms} we learned that either needing too many modes to model the ionospheric correction or too many orders of correction leads to an unreasonable number of numerical calculations.  Second, from section \ref{nmax} we learned that under correcting in the $uv$-plane is a huge mistake and a lot worse than not correcting at all.  Hence the dilemma:  choosing $n_{max}$ too small leads to the destruction of the data, while choosing $n_{max}$ too large leads to a computationally infeasible problem.  It is therefore advantageous to develop a theoretical prediction of how many orders of correction are necessary.  As it turns out, the result will lead to a few tricks which make the problem more reasonable.

\subsection{Finding an Upper Bound on the Error $U(\vec{u})$, and $n_{max}$}
\hspace{4 mm}  If we correct to only $n_{max}$ orders, then the magnitude error $E(\vec{u})$ in our result must be the absolute value of the sum of all the terms we left out; more specifically,
{\footnotesize
\begin{equation}
E(\vec{u}) = \left| \sum_{n=n_{max} + 1}^{\infty} \sum_{l_1, l_2, ... , l_{2M}}' \left( \prod_{q=1}^{2M} \frac{ ( c_q \vec{u} \cdot \vec{d}_q)^{l_q}}{l_q!} \right) \tilde{I}(\vec{u} + \sum_q l_q \vec{d}_q ) \right| .
\end{equation}
} 
The steps leading to an upper bound $U(\vec{u})$ on this error may be found in the appendix, section \ref{math2}.  The result is
{\footnotesize
\begin{equation}
\label{upper}
U(\vec{u}) = \sum_{n=n_{max} + 1}^{\infty} I_{max}   v(\vec{u}) ^{n} \frac{(n + 2M - 1)!}{n!(2M-1)!} \frac{1}{\left( (\frac{n}{2M})! \right)^{2M} } \geq E(\vec{u}),
\end{equation}
} 
where 
\begin{equation}
\label{vu}
v(\vec{u}) = \mbox{\hspace{1 mm} max \{  } | c_q \vec{u} \cdot \vec{d}_q | \hspace{1 mm} \}  \mbox{\hspace{5 mm} any \hspace{1 mm}  } q, 
\end{equation}
\begin{equation}
I_{max} = \mbox{\hspace{1 mm} max \{  } |\tilde{I}(\vec{u})| \hspace{1 mm} \}  \mbox{\hspace{5 mm} any \hspace{1 mm}  } \vec{u}, 
\end{equation}
and $M$ is the number of modes, as always.  [See equations \ref{c_qdef} and \ref{d_qdef} for reminders on how effective modes $(c_q, \vec{d}_q)$ are related to actual modes $(a_m, \vec{b}_m)$].  This is our final result for a strict upper bound on the error.  Unfortunately, this formula is not too enlightening.  In order to obtain a theoretical estimate for $n_{max}$, we must make a few further approximations.  

\begin{figure}
\begin{center}
\PSbox{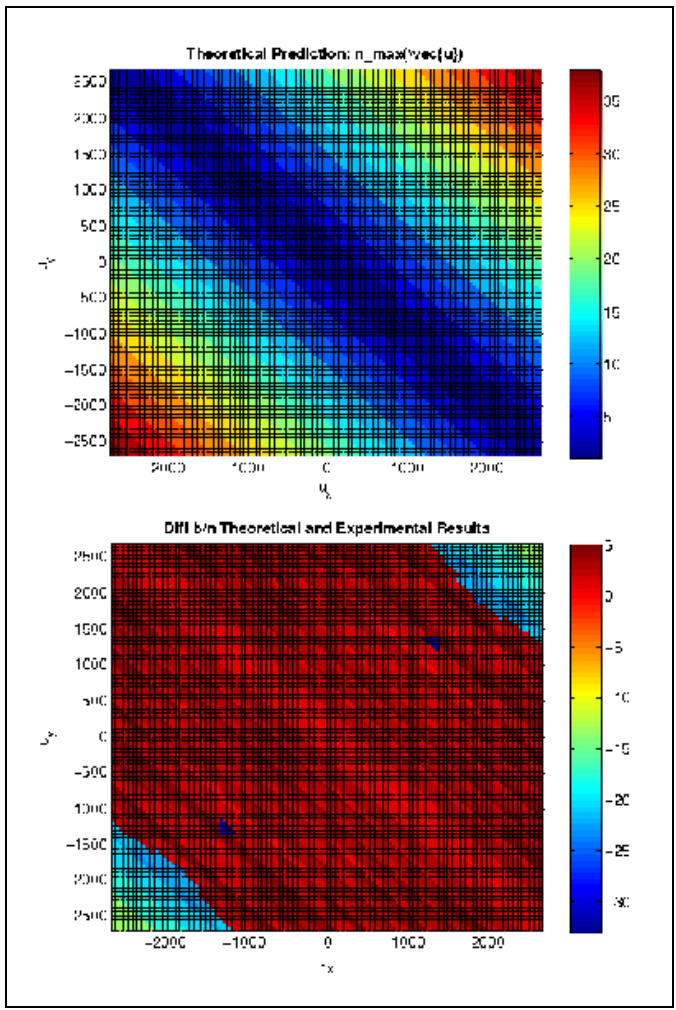}{3.4in}{5.15in}
\caption{Theoretical Predictions for $n_{max}$. \emph{Top Panel}  The theoretical predictions for $n_{max}$ shown graphically as a function of $\vec{u}$ for the simple sky explored in section \ref{nmax} (and shown in figure \ref{pureandbad}).  This plot has $P=.1$ (see equations \ref{P} and \ref{finalnmax}) .  The axes are measured in inverse radians, and the third dimension with color represents the calculated value of $n_{max}$.  The preferred direction is defined by the direction of the ionospheric perturbation ($\pi/4$ with respect to the $u_x$ axis).  Notice that the value of $n_{max}$ increases as $\vec{u}$ moves radially outward along this direction.  \emph{Bottom Panel}  The residual between the theoretical prediction shown in the top panel and the numerical calculations of $n_{max}$ for the same sky.  For a bulk of the $uv$-plane, the prediction is quite good (within 5).  The cool colored areas represent pixels that never obtain the desired fractional error due to the non-locality of the correction (see section \ref{trunc10}).     }
\label{nmaxtheory}
\end{center}
\end{figure}

\begin{figure}
\begin{center}
\PSbox{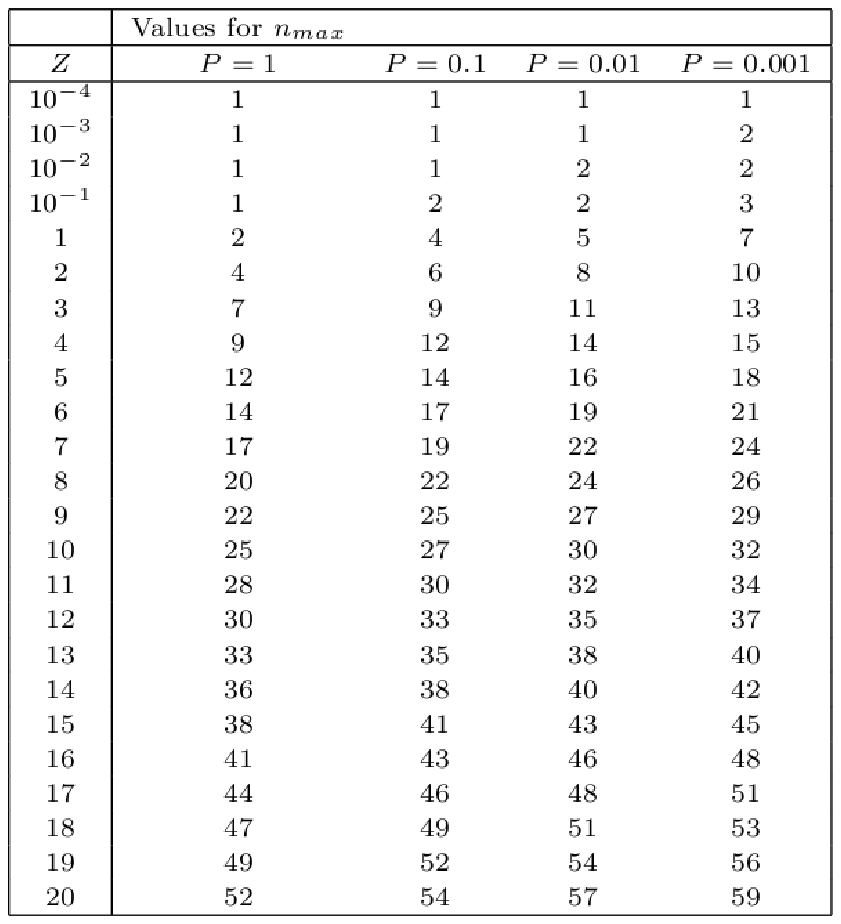}{3.4in}{3.75in}
\label{nmaxtab}
\end{center}
\end{figure}

As one would expect, the optimal value of $n_{max}$, which represents the number of orders necessary to obtain some level of accuracy in the $uv$-plane, is dependent upon the level of accuracy desired.  To quantify this, define $I_{n_{max}}$ to be the value of the intensity in the $uv$-plane after correcting up through $n = n_{max}$, and $I_{actual}$ to be the value of the intensity in the $uv$-plane that one would obtain by employing the full correction and not truncating the sum (ie, $n_{max} = \infty$).  (It should be noted that $I_{actual}$ here also assumes a $uv$-plane infinite in extent.  This will have effects seen later.)  The fractional error $f$ in the $uv$-plane correction caused by truncating the sum is then 
\begin{equation}
\label{P}
f = \frac{| I_{n_{max}} - I_{actual} |}{|I_{actual}|} .
\end{equation}
In the appendix (section \ref{math3}) you will find the steps leading up to a theoretical prediction of the value of $n_{max}$ at a given $\vec{u}$ in the $uv$-plane necessary to obtain a fractional error less than or equal to $P$ if given the ionospheric effective modes distorting the sky, $(c_q, \vec{d}_q)$.  The result is that the optimal value of $n_{max}$ is estimated by
\begin{equation}
\label{finalnmax}
n_{max} = \mbox{\hspace{1 mm} min \{  } n \mbox{ \} \hspace{1 mm} such that \hspace{1 mm} } 
Pn! \geq (2Mv(\vec{u}))^n,
\end{equation}
where $M$ is the number of modes and $v(\vec{u})$ is defined as it was above in equation \ref{vu}. This formula is a little hard to digest, so some values for $n_{max}$ given $Z \equiv 2Mv(\vec{u})$ and $P$ are provided in table \ref{nmaxtab}.  For the example sky and perturbation used in section \ref{nmax}, the theoretical predictions for the number of orders necessary is shown graphically in the top panel of figure \ref{nmaxtheory}.  This figure has $P = .1$, although $n_{max}$ does not change too significantly when varying $P$, as is seen in table \ref{nmaxtab}.  Recall that the single distorting mode $\vec{d}_q$ is in the $\pi/4$ direction, which defines the favored direction seen in this figure.

As a check of these theoretical predictions, we used MATLAB to numerically compute the number of orders necessary to obtain the desired fractional error of $P=.1$.  The bottom panel of figure \ref{nmaxtheory} shows the difference between these computational results and the theoretical predictions shown in the top panel of figure \ref{nmaxtheory}.  More specifically, it represents the number of orders of correction theoretically predicted minus the number found computationally.  This figure suggests that for a bulk of the $uv$-plane, the theoretical prediction is quite accurate, predicting the number of orders to within 5.  Near the extremes, however, the finite nature of the $uv$-plane causes problems (remember that the theoretical estimate assumed an infinite $uv$-plane).  In fact, the cool-colored pixels near the corners are pixels which never obtained a fractional error of $P < .1$ (The numerical code cutoff after 50 orders; all points with fractional errors too high at that point were assigned a value of 50 orders).

The theory predicts that about 35 orders are required to correct at the most extreme points in the $uv$-plane, which is what our previous numerical computations found.  An important feature of the theoretical predictions shown in figure \ref{nmaxtheory} that is characteristic of all skies is that the necessary number of orders of correction varies with $\vec{u}$, and in particular it increases as $\vec{u}$ increases along the direction of the mode.  Therefore, points closer to the origin are corrected in less orders than those further away.  

\subsection{The Strongest Mode and Significant Modes Approximations}
\label{strongest}

\hspace{4 mm}  The accuracy of the theoretical prediction here is in no small part due to the existence of only one ionospheric mode in our simple sky model.  This reason for this is that the above theoretical estimate (equation \ref{finalnmax}) was derived from an expression for the upperbound on the error (equation \ref{upper}) which assumes that all the ionospheric modes in the sky are as strong as the strongest mode at $\vec{u}$ and add constructively (which may be seen in the appendix, section \ref{math2}, near equation \ref{Uu}).  As such, the result is not a bad prediction for only one distorting mode, but tends to (perhaps significantly) overestimate the necessary number of orders for multiple distorting modes.  In short, the above mentioned theoretical estimate may perhaps be more accurately called a theoretical \emph{over}estimate.  Given the results seen in section \ref{nmax} (more specifically, the terrible consequences of undercorrecting in the $uv$-plane), this was done intentionally to ensure that our $uv$-plane was adequately corrected.  Still, it may be useful to obtain a more accurate estimate of the number of orders necessary.

One such estimate would be a \emph{strongest mode approximation}, in which we assume that at any given $\vec{u}$, the only significant contribution comes from the strongest mode at that point.  The contributions from the other modes are assumed to be weak and negligible.  This approximation ultimately boils down to setting $M=1$ in the final equation determining $n_{max}$ from the previous section (equation \ref{finalnmax}).  This approximation may provide a more accurate estimate of $n_{max}$, but it also runs a high risk of underestimating the correct number of orders, which should be avoided if possible.  An alternate approximation would be a \emph{significant modes approximation}, in which only modes at a given $\vec{u}$ with strengths within a certain critical fraction of the strongest mode's are included in the value for $M$ used in equation \ref{finalnmax}.  With a closer study of perturbations from more realistic ionospheric modes, it may be possible to set this critical fraction in such a way as to fairly accurately predict the number of orders necessary.

\section{Two Methods for Making the $uv$-plane Correction More Feasible}
\label{feasible}

\hspace{4 mm}  The above analysis suggests two methods for making the $uv$-plane correction less time intensive:  1) The points furthest out in the $uv$-plane take the most time to correct.  Eliminating them decreases computation time, but at the cost of resolution in the real space sky.  2)  Different points in the $uv$-plane require different numbers of orders of correction, so write a code that corrects to different numbers of orders depending on the point in the $uv$-plane.  (In other words, don't waste time correcting to 35 orders near the origin when 2 is enough.)

\begin{figure}
\begin{center}
\PSbox{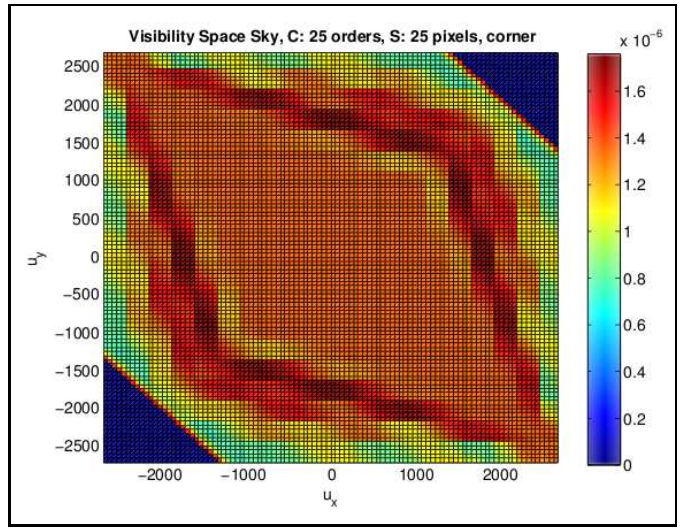}{3.4in}{2.64in}
\caption{\emph{The Edge Shaving Technique}  This is a visual example of the edge shaving techinique, in which the pixels in the 25 diagonal rows from the upper right and lower left corners (ie, the pixels which require the most orders of correction) have had their values set to zero (or, are \emph{shaved}).  Notice that eliminating these ultra intense pixels reveals some of the finer patterns caused by the $uv$-plane correction, as also seen in the correction to 35 orders (figure \ref{order35}, bottom panel).  Compare this to the $uv$-plane created by correcting to 25 orders and not edge shaving (figure \ref{order25}, bottom panel), in which the intense pixels from the corner dominate the others in the $uv$-plane. }
\label{uvshave}
\end{center}
\end{figure}

\begin{figure}
\begin{center}
\PSbox{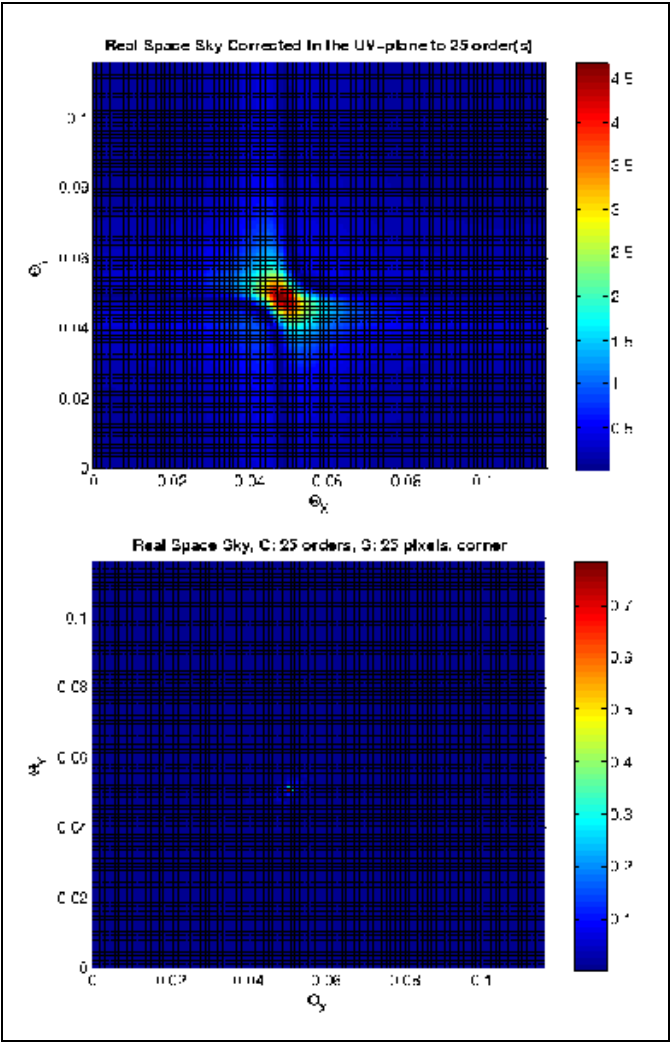}{3.4in}{5.35in}
\caption{The results of the edge shaving technique.  \emph{Top Panel}  A reproduction of the top panel of figure \ref{order25}, shown as a reminder of the real space sky that results by correcting to 25 orders in the $uv$-plane and not edge shaving. \emph{Bottom Panel}  The real space sky that results from correcting to 25 orders and shaving 25 rows of pixels from the corner in visibility space (ie, the real space sky created by inverse Fourier transforming the $uv$-plane shown in figure \ref{uvshave}).  Notice the drastic improvement:  The sky now looks like a single source of approximately the right intensity and at the right location, but with a small spreading of intensity to neighboring pixels.  }
\label{corner25}
\end{center}
\end{figure}

\begin{figure}
\begin{center}
\PSbox{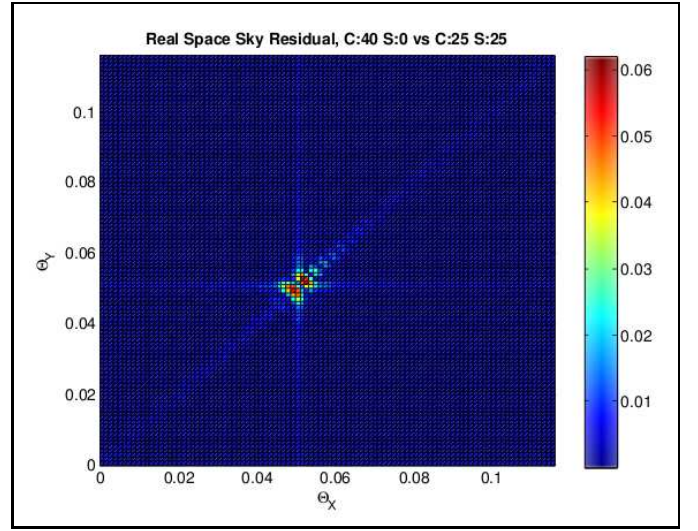}{3.4in}{2.66in}
\caption{This is the residual between the real space sky created by fully correcting to 40 orders with no edge shaving and the real space sky created by fully correcting to 25 orders and then edge shaving 25 rows of pixels (shown in the bottom panel of figure \ref{corner25}).  Notice that the two skies seem to match pretty well, with with maximum disparities an order of magnitude lower than the maximum intensity of the source.  }
\label{residshave}
\end{center}
\end{figure}

\begin{figure}
\begin{center}
\PSbox{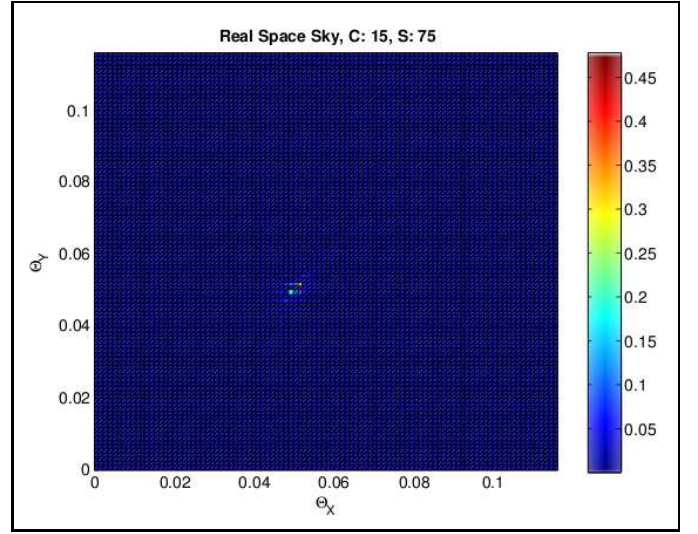}{3.4in}{2.68in}
\caption{The real space sky that results from correcting to only 15 orders, and then shaving 75 diagonal rows of pixels in the $uv$-plane.  Comparison of this real space sky to that obtained by correcting to 15 orders and not edge shaving (see the top panel of figure \ref{order15}) shows remarkable improvement.  The maximum intensity has decreased by a factor of 1000, and is now of the right order of magnitude.  On the down side, however, the source is now spread among a fair number of pixels, and might not posses the precision one desires.       }
\label{realshave15}
\end{center}
\end{figure}
\subsection{Method 1: Eliminating the Extremes of the $uv$-plane}
\label{meth1}

\hspace{4 mm}  For this method, we eliminate the problems caused by under correcting at the extremes in the $uv$-plane by setting the values at those extremes to 0.  Take, as a visual example, figure \ref{uvshave}, in which we have set the values of the pixels in the 25 diagonal rows from the corner to zero (we \emph{shaved} 25 pixels from the corner).  As a reminder of the real space sky after 25 orders without edge shaving, consider the top panel of figure \ref{corner25} (a reproduction of figure \ref{order25}, top panel).  Notice that the intensity is approximately 4 or 5 times too high at the brightest points and, even worse, our single point source has turned into some sort of supernova explosion.  Compare this to the edge-shaved version of the real space sky (figure \ref{corner25}, bottom panel), in which we see a sky that looks almost identical to our fully corrected sky after 35 orders (top panel, figure \ref{order35}).  To see this more clearly, consider figure \ref{residshave}, which represents (the absolute value of) the residual between the real space sky corrected to 40 orders with no edge shaving and the real space sky corrected to 25 orders with edge shaving.  From this figure we see that the result of the edge shaving was to create a small spread around the star, but of an intensity about an order of magnitude lower than the maximum intensity of the star.

The relative success of this scheme leads to the question of how low we may push the number of correction orders when edge shaving is introduced.  Figure \ref{realshave15} shows the result of only correcting to 15 orders, but shaving 75 rows of pixels from the corners in the $uv$-plane.  Without edge shaving, the real space sky corrected to 15 orders had a maximum intensity of about 500 (figure \ref{order15}, top panel).  Now, the total intensity is approximately 1 as it should be, but it is spread over a number of pixels.  So while the results are a dramatic improvement over what they had been, for the sake of precision it might be a good idea to correct to higher orders and shave less.  The moral:  The process works, but be careful about trying to shave too much.

\subsection{Method 2: Correcting to Different Orders at Different $\vec{u}$}
\label{meth2}

\begin{figure}
\begin{center}
\PSbox{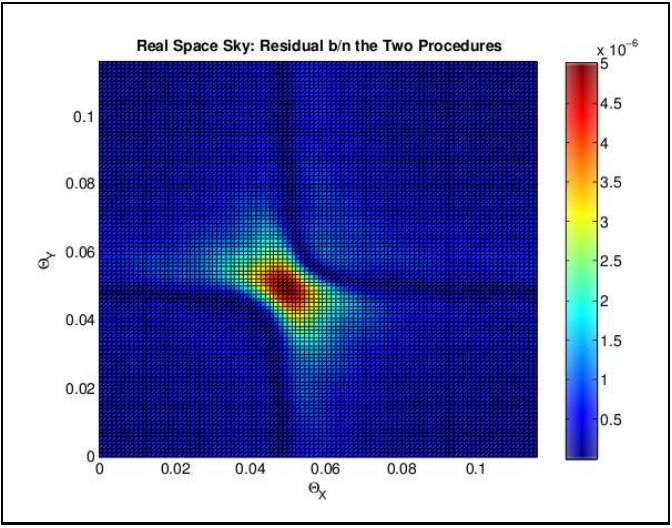}{3.4in}{2.64in}
\caption{The residual between the real space sky created by correcting to 40 orders at all points in the $uv$-plane and the real space created by correcting to varying orders in the $uv$-plane, as determined by the theoretical prediction for $n_{max}$ (equation \ref{finalnmax}; shown graphically for the simple sky of section \ref{nmax}, which was also used here, in the top panel of figure \ref{nmaxtheory}).  Notice that the maximum difference is 5 or 6 orders of magnitude less than the maximum in the intensity (Recall that the correction to 40 orders is essentially the same as that to 35 orders, which is shown in the top panel of figure \ref{order35}).  This suggests once again that the theoretical prediction for the number of orders is fairly accurate.    }
\label{residmeth2}
\end{center}
\end{figure}

\hspace{4 mm}  For this method, we correct to different numbers of orders at different $\vec{u}$.  To test this method, we re-wrote our MATLAB code so that the number of orders of correction at a given $\vec{u}$ was determined by the theoretical estimate from section \ref{theorynmax} (More specifically, equation \ref{finalnmax}).  We then corrected the distortion for the same simple sky used in section \ref{nmax}.  Figure \ref{residmeth2} shows the residual between the real space sky corrected to 40 orders at all points in the $uv$-plane and the real space sky corrected to different orders in the $uv$-plane.  Not surprisingly, the residual is incredibly small-- 5 or 6 orders of magnitude less than the maximum in the intensity of the source.  (This is, of course, another sign that the theoretical prediction of the number of orders of correction is pretty good).  However, this new MATLAB code presented a small problem:  MATLAB is so much better at manipulating matrices than running for-loops that this second code, which theoretically requires less computation, takes approximately 15 times as long to run.  Of course, if the $uv$-plane correction is eventually used in MWA, a programming language more adept at loops will undoubtfully be used, and this method will potentially save time.

\section{An Alternate Choice for $\vec{\delta \theta}'$}
\label{choice2}

\hspace{4 mm} Previously, we have assumed that $\vec{\delta \theta}'$ took the form

\begin{equation}
\delta \vec{\theta}' = Re( \sum_{m=1}^{M} ia_m \vec{b}_m e^{-i \vec{b}_m \cdot \vec{\theta}' }  ).
\end{equation}
Given the relative complexity of the results above, it pays to investigate an alternate choice of $\vec{\delta \theta}'$.  The main reason that this was chosen was because it allowed for an analytic solution for $I(\vec{u})$, where
\begin{equation}
\label{second}
I(\vec{u}) = \int \frac{d^2u'}{(2\pi)^2} \int d^2 \theta' e^{i \vec{\theta}' \cdot (\vec{u}' - \vec{u}) } e^{ -i \vec{u} \cdot \delta \vec{\theta}' } \tilde{I}(\vec{u}') , 
\end{equation}
as shown previously in equation \ref{IUpure}.  Without an analytic solution to these integrals for $I(\vec{u})$, the $uv$-plane correction becomes more computationally intensive than inverse Fourier transforming to real space, correcting for the ionosphere there, and then Fourier transforming back to visibility space.  Therefore, an analytic solution is required for any choice suitable choice of $\vec{\delta \theta}'$.  Unfortunately, choices for $\vec{\delta \theta}'$ which allow such analytic solutions are hard to find.  There is, however, at least one other such distortion:  a polynomial expansion, given by 
\begin{equation}
\delta \vec{\theta}' = \sum_{k=1}^{K} a_k \theta_x'^{k} + \sum_{m=1}^{M} b_m \theta_y'^{m} , 
\end{equation}
where $K$ and $M$ are the number of terms in the $\theta_x$ and $\theta_y$ directions, respectively, necessary to accurately model the distortion by the ionosphere.  The $a_k$ and $b_m$ shown here are real.  The analytic derivation of $I(\vec{u})$ is given in the appendix, section \ref{math4}.  The result is
{ \footnotesize
\begin{eqnarray}
I(\vec{u}) = \sum_{n_x}^{\infty} & & \sum_{n_y}^{\infty} \sum_{l_m}' \sum_{l_k}' - (-i u_x)^{n_x-t_x+1} (-i u_y)^{n_y-t_y+1} \\
 & & \mbox{ {\textbf X } } \prod_{m=1}^{M} \frac{(b_m)^{l_m}}{l_m!} \prod_{k=1}^{K} \frac{(a_k)^{l_k}}{l_k!} \frac{\partial^{t_x}}{\partial u_x^{t_x}} \frac{\partial^{t_y}}{\partial u_y^{t_y}} \tilde{I}(\vec{u}) 
\end{eqnarray}
} 
where
\begin{equation}
t_y = \sum_m m l_m , 
\end{equation}
\begin{equation}
t_x = \sum_k k l_k ,
\end{equation}
and the sums over $l_m$ and $l_k$ are restricted so that $\sum_k l_k = n_x$ and $\sum_m l_m = n_y$.

While we've confined the analytic solution of this to the appendix, it should be mentioned that en route to this solution the exponential in $\vec{\delta \theta}'$ from equation \ref{second} was Taylor-expanded (which, you may recall, was also the case for the other choice of $\vec{\delta \theta}'$ as a sum over sinusoidal modes, and is explicitly shown in equations \ref{n1} and \ref{n2}).  In other words, this solution is likewise characterized by the double expansion mentioned previously at the end of section \ref{sinmodes}: one expansion over $(n_x, n_y)$ from expanding the exponential in $\vec{\delta \theta}'$, and one expansion in $(l_k, l_m)$ resulting from our model for the ionosphere.  As such, this solution shows many of the unfortunate characteristics of the sinusoidal choice.  In particular, there are still restricted sums which contain a number of terms comparable to that calculated in section \ref{numterms}, so this choice has the same problem of making the $uv$-plane correction unreasonable if too many orders of correction or ionospheric modes are needed.

In addition, the correction is still not strictly local:  numerical computation of derivatives requires neighboring pixels, with higher orders requiring more neighbors.  Moreover, numerical computation of derivatives for a finite data set introduces its own set of additional errors, and thus makes this choice much less appealing than the previous one with sinusoidal modes.

The ugliness of both of these choices ultimately stems from the inability to solve for $I(\vec{u})$ analytically without expanding the exponential containing $\vec{\delta \theta}'$ .  Unless a choice is found which may be solved analytically without this first expansion, it is doubtful that a better choice than the sinusoidal modes will be found.

\section{A More Realistic Ionosphere}
\label{realistic}

\hspace{4 mm} The ionospheric distortion presented throughout this paper was chosen because its particularly strong nature accentuated the subtleties of the $uv$-plane correction.  The strength of this mode made the $uv$-plane correction appear computationally infeasible: any ionosphere which required even 10 of these modes to accurately model would require too much computation (see section \ref{numterms}).  However, an ionospheric mode which shifts sources on the sky by tens of arcmin is somewhat unrealistic.  We conclude by considering a more realistic distortion.

The computational feasibility of the $uv$-plane correction is determined by the largest value of $n_{max}$ (given by equation \ref{finalnmax}) required for any $\vec{u}$.  To calculate this for a realistic sky, we need to know the largest possible value of $v(\vec{u})$.  It is possible to cast the largest value of $v(\vec{u})$, which we label $v_{max}$, in a form which better elucidates its physical significance.  More specifically, notice that
\begin{eqnarray}
v(\vec{u}) & = & \mbox{\hspace{1 mm} max \{  } | c_q \vec{u} \cdot \vec{d}_q | \hspace{1 mm} \}  \mbox{\hspace{5 mm} any \hspace{1 mm}  } q \\
& = & \mbox{\hspace{1 mm} max \{  } | c_q |\vec{u}| |\vec{d}_q| \cos(\alpha) | \hspace{1 mm} \}  \mbox{\hspace{5 mm} any \hspace{1 mm}  } q,
\end{eqnarray}
where $\alpha$ is the angle between $\vec{u}$ and $\vec{d}_q$.  For an arbitrary choice of $\vec{u}$, this cosine term may be significant.  However, if we wish to calculate $v_{max}$, we may set $\alpha = 0$ and $|\vec{u}|$ = $u_{max}$, where $u_{max}$ is the greatest distance from the origin in the $uv$-plane that our antenna's $uv$-coverage allows.  This is valid for MWA because the $uv$-coverage is approximately circular, so that the strongest ionospheric distorting mode is guaranteed to lie along a direction which possesses this maximum displacement in the $uv$-plane.  With these changes, and based upon our previous definitions of the effective modes $\{ c_q, \vec{d}_q \}$ in terms of the actual modes $\{ a_m, \vec{b}_m \}$ (see equations \ref{c_qdef} and \ref{d_qdef}), we may write $v_{max}$ as
\begin{equation}
v_{max} = \mbox{\hspace{1 mm} max \{  } | \frac{1}{2} a_m \vec{b}_m | u_{max} \hspace{1 mm} \}  \mbox{\hspace{5 mm} any \hspace{1 mm}  } m .
\end{equation}
But,
\begin{eqnarray}
|a_m \vec{b}_m | & = & |i a_m \vec{b}_m e^{-i \vec{b}_m \cdot \vec{\theta}'}| \\
& \geq & | Re(i a_m \vec{b}_m e^{-i \vec{b}_m \cdot \vec{\theta}'} )| = |\delta \vec{\theta}'|
\end{eqnarray}
Let's define
\begin{equation}
| \delta \vec{\theta}_{max} | \equiv \mbox{\hspace{1 mm} max \{  } | a_m \vec{b}_m | \hspace{1 mm} \}  \mbox{\hspace{5 mm} any \hspace{1 mm}  } m .
\end{equation}
In words, $| \delta \vec{\theta}_{max} |$ is the maximum deflection caused by a single mode that we might observe.  In addition, let $B$ denote the length of our antenna array's longest baseline.  The maximum $uv$-plane displacement $u_{max}$ is then this length measured in units of the wavelength $\lambda$ that our antenna is detecting,
\begin{equation}
u_{max} = \frac{2\pi B}{\lambda} . 
\end{equation}
[The extra factor of $2\pi$ is the result of our convention for Fourier transforms (see equation \ref{Fourierdef}), which differs from that conventionally used in radio astronomy].  With these substitutions,
\begin{equation}
v_{max} = \frac{1}{2} | \delta \vec{\theta}_{max} | \frac{2\pi B}{\lambda} .
\end{equation}
Therefore, in terms of these parameters, the number of orders of correction necessary is (adapted from equation \ref{finalnmax})
{\small
\begin{equation}
n_{max} = \mbox{\hspace{1 mm} min \{  } n \mbox{ \} \hspace{1 mm} such that \hspace{1 mm} } 
Pn! \geq (M  | \delta \vec{\theta}_{max} | \frac{2\pi B}{\lambda} )^n,
\end{equation}
}
where $P$ is (as before) the fractional error desired for the correction.  It should be noted that this form is only valid for determining the largest $n_{max}$ among all $\vec{u}$.  For calculating $n_{max}$ for a particular $\vec{u}$, equation \ref{finalnmax} must be used.

For MWA, a typical frequency detected will be about 140 MHz, corresponding to $\lambda \approx 2.143$ meters (this represents the 21 cm emission for a red shift of $z \approx 9.2$.)  We expect the ionosphere to deflect such a wave approximately 0.6 arcmin = $1.75*10^{-4}$ radians (\cite{D}, value is for the night).  If we consider baselines of approximately 400 meters, then
\begin{equation}
Z_{full} \equiv M  | \delta \vec{\theta}_{max} | \frac{2\pi B}{\lambda} \approx .205 M .  
\end{equation}
If given the number of modes necessary to accuately model the ionosphere $M$ (which is as of yet undetermined), then $n_{max}$ for the full array may be determined from table \ref{nmaxtab} by substituting $Z_{full}$ for $Z$.  As an example, if a fractional error of $P = .1$ is desired and $M = 20$, then $Z_{full} \approx 4$ and table \ref{nmaxtab} shows that 12 orders of correction are necessary.  Whether such a result is computationally feasible is dependent upon how much time is alloted for the ionospheric correction and the quality of the computers used.  As such, it may not be determined here.  What is clear, however, is that such a correction is not obviously ruled out on computational grounds (especially if a technique such as edge shaving is used to reduce $n_{max}$). [Quick aside:  Edge shaving alters the above results by substituting the largest $|\vec{u}|$ left unshaved in place of $u_{max}$ in the above calculations.]  It should be noted that the values that went into calculating $Z_{full}$ above were estimates, and certainly not set in stone.  In particular, we once again emphasise that throughout this paper we have remained ignorant of the details involved in the expansion of the ionosphere, and have no knowledge of how many modes $M$ are necessary to sufficiently model the effect of the ionosphere.  In addition, the ionospheric deflection $ | \delta \vec{\theta}_{max} | $ is proportional to $\lambda^2$, with longer wavelengths experiencing greater shifts \cite{TMS}.  Therefore, we expect longer wavelengths than the above to require more orders of correction and shorter wavelengths, fewer.  If $Z_{full}$ is in fact lower by a factor of 5, for example, then $n_{max} \approx 4$ and the $uv$-plane correction is certainly a viable candidate for correcting the ionosphere.  In particular, if the strongest mode approximation discussed in section \ref{strongest} turns out to be a good approximation, then even with baselines of 1.5 km we may expect a good correction after only 4 orders for the wavelength given above.  On the other hand, if $Z_{full}$ is raised by a factor of 5, then $n_{max} \approx 54$ and the $uv$-plane correction is clearly computationally infeasible for any reasonable value of $M$.

\section{Potential Computation Saver:  Updating the {\bf A$^{T}$} matrix}

\hspace{4 mm}  As stated previously, correcting for the ionosphere in the $uv$-plane entails multiplying the perturbed data $\tilde{I}(\vec{u}')$ by the ionospheric correction operator {\bf A$^{T}$}$(\vec{u},\vec{u}')$,
\begin{equation}
I(\vec{u}; t) = {\bf A^{T}}(\vec{u},\vec{u}'; t) \tilde{I}(\vec{u}'; t) ,
\end{equation}
where we have now made the time dependence of these quantities explicit.  The above may be thought of as a matrix equation, where $\tilde{I}(\vec{u}';t)$ and $I(\vec{u};t)$ represent our uncorrected and corrected (respectively) data arrays, and ${\bf A^{T}}(\vec{u},\vec{u}';t)$ represents a correction matrix.  The entries of this correction matrix are calculated by the appropriate binning of the coefficients in our correction equation (reproduced from \ref{finalIU}),
{\small
\begin{equation}
I(\vec{u};t) = \sum_{n=0}^{n_{max}} \sum_{l_1, l_2, ... , l_{2M}}' \left( \prod_{q=1}^{2M} \frac{ ( c_q \vec{u} \cdot \vec{d}_q)^{l_q}}{l_q!} \right) \tilde{I}(\vec{u} + \sum_q l_q \vec{d}_q; t) ,
\end{equation}
} 
where the ``coefficients" are the quantities preceeding $\tilde{I}(\vec{u} + \sum_q l_q \vec{d}_q; t)$ on the right hand side of the equation.  If the timescale within which one wishes to recalculate the effect of the ionosphere is small compared to the timescale within which the ionosphere significantly changes, then it is possible that the correction matrix {\bf A$^{T}$}$(\vec{u},\vec{u}';t)$ has changed very little from that previously calculated.  More specifically, if one wishes to calculate the correction matrix at a time $t_1$ shortly after having calculated it at time $t_0$ (ie, if $|t_1 - t_0| << \tau$ where $\tau$ is the time scale of significant change in the ionosphere), then
\begin{equation}
{\bf A^{T}}(\vec{u},\vec{u}'; t_1) \approx {\bf A^{T}}(\vec{u},\vec{u}'; t_0) + \delta {\bf A^{T}}(\vec{u},\vec{u}'; t_0)  
\end{equation}
where $\delta {\bf A^{T}}$ represents a small correction matrix.  In this regime, it is computationally much easier to calculate the small correction $\delta {\bf A^{T}}$ and add it to the previously calculated ${\bf A^{T}}(\vec{u},\vec{u}'; t_0)$ then to calculate  ${\bf A^{T}}(\vec{u},\vec{u}'; t_1)$ from scratch.  Therefore, in such a scenario the process of updating the correction matrix is computationally favorable.

The first order correction $\delta {\bf A^{T}}$ may be analytically calculated as follows:  Let's assume that we model the ionosphere using the effective modes ${c_q, \vec{d}_q}$ that we have been throughout this paper (see equations \ref{c_qdef} and \ref{d_qdef}), and that the $\vec{d}_q$ represent part of a fixed Fourier basis while the $c_q$ are our fitting parameters.  (In other words, the $\vec{d}_q$ are fixed and time independent, while the $c_q$ fluctuate with time).  Let's assume that we've calculated the correction at a time $t_0$.  More specifically, assume that for all $\vec{u}$ we've calculated and stored all the relevent terms in the correction equation for $I(\vec{u}; t_0)$,
\begin{eqnarray}
\label{t_0}
I(\vec{u}; t_0) = \sum_{n=0}^{n_{max}} & & \sum_{l_1, l_2, ... , l_{2M}}' \left( \prod_{q=1}^{2M} \frac{ ( c_q(t_0) \vec{u} \cdot \vec{d}_q)^{l_q}}{l_q!} \right) \\
& & \mbox{ {\textbf X } \hspace{2 mm} } \tilde{I}(\vec{u} + \sum_q l_q \vec{d}_q; t_0 ) ,
\end{eqnarray}
where the time dependence of $I$, $\tilde{I}$, and $c_q$ is now explicit.  Now we want to correct for the ionosphere at a later time $t_1$.  To compose the correction matrix at time $t_1$, we could start with the full correction formula,
\begin{eqnarray}
I(\vec{u}; t_1) = \sum_{n=0}^{n_{max}} & & \sum_{l_1, l_2, ... , l_{2M}}' \left( \prod_{q=1}^{2M} \frac{ ( c_q(t_1) \vec{u} \cdot \vec{d}_q)^{l_q}}{l_q!} \right) \\
& &  \mbox{ {\textbf X } \hspace{2 mm} } \tilde{I}(\vec{u} + \sum_q l_q \vec{d}_q; t_1 ) ,
\end{eqnarray}
and then construct the new matrix ${\bf A^{T}}(\vec{u},\vec{u}'; t_1)$.  Instead, however, let's assume that we're in the regime of small ionospheric changes, so that to first order
\begin{equation}
c_q(t_1) \approx c_q(t_0) + \Delta c_q,
\end{equation}
where $\Delta c_q$ is small.  Substituting this into the full correction and only keeping terms to first order we obtain
{ \footnotesize
\begin{eqnarray*}
I(\vec{u}, t_1) = & & \sum_{n=0}^{n_{max}} \sum_{l_1, l_2, ... , l_{2M}}' \left( \prod_{q=1}^{2M} \frac{ ( (c_q(t_0)+\Delta c_q) \vec{u} \cdot \vec{d}_q)^{l_q}}{l_q!} \right) \\
& & \mbox{ \hspace{6 mm} {\textbf X } \hspace{2 mm} } \tilde{I}(\vec{u} + \sum_q l_q \vec{d}_q ; t_1 ) \\
 = & & \sum_{n=0}^{n_{max}} \sum_{l_1, l_2, ... , l_{2M}}' \left( \prod_{q=1}^{2M} \frac{ ( c_q(t_0) \vec{u} \cdot \vec{d}_q)^{l_q}(1+l_q \frac{\Delta c_q}{c_q(t_0)})}{l_q!} \right) \\
& & \mbox{ \hspace{6 mm} {\textbf X } \hspace{2 mm} } \tilde{I}(\vec{u} + \sum_q l_q \vec{d}_q ; t_1 ) \\
= & & \sum_{n=0}^{n_{max}} \sum_{l_1, l_2, ... , l_{2M}}' \left( \prod_{q=1}^{2M} \frac{ ( c_q(t_0)\vec{u} \cdot \vec{d}_q)^{l_q}}{l_q!} \right) \left( 1 + \sum_{q=1}^{2M} l_q \frac{\Delta c_q}{c_q(t_0)}  \right) \\
& & \mbox{ \hspace{6 mm} {\textbf X } \hspace{2 mm} }  \tilde{I}(\vec{u} + \sum_q l_q \vec{d}_q; t_1  ) \\
= & & \sum_{n=0}^{n_{max}} \sum_{l_1, l_2, ... , l_{2M}}' \left( \prod_{q=1}^{2M} \frac{ ( c_q(t_0)\vec{u} \cdot \vec{d}_q)^{l_q}}{l_q!} \right) \tilde{I}(\vec{u} + \sum_q l_q \vec{d}_q; t_1  ) + \\
& & + \sum_{n=0}^{n'_{max}} \sum_{l_1, l_2, ... , l_{2M}}' \left( \prod_{q=1}^{2M} \frac{ ( c_q(t_0)\vec{u} \cdot \vec{d}_q)^{l_q}}{l_q!} \right) \left( \sum_{q=1}^{2M} l_q \frac{\Delta c_q}{c_q(t_0)}  \right) \\
& & \mbox{ \hspace{6 mm} {\textbf X } \hspace{2 mm} } \tilde{I}(\vec{u} + \sum_q l_q \vec{d}_q ; t_1 ) .
\end{eqnarray*}
}
In this final equation, the coefficients in this first group of terms exactly replicate those coefficients from time $t_0$ (see equation \ref{t_0}).  In other words, these terms represent the previously determined correction matrix ${\bf A^{T}}(\vec{u},\vec{u}'; t_0)$.  The second group represents the small adjustment $\delta {\bf A^{T}}$ to the correction matrix at $t_0$.  Notice that these terms have been there own maximum cutoff for $n$, labeled as $n'_{max}$ in the above equation.  If the quantity  $\Delta c_q/c_q(t_0)$ is small, as assumed, then the individual terms in this second sum are also small, and thus a smaller value of $n'_{max}$ is necessary to obtain a desired fractional error for the intensity.  In this case, it is computationally favorable to update the correction matrix rather than derive it from scratch.  It should also be noted that if $n'_{max} < n_{max}$, then the nonzero entries of the matrix $\delta {\bf A^{T}}$ form a subset of the nonzero entries of the matrix ${\bf A^{T}}(\vec{u},\vec{u}'; t_0)$, and thus their sum (which represents ${\bf A^{T}}(\vec{u},\vec{u}'; t_1)$) is equally as sparse as ${\bf A^{T}}(\vec{u},\vec{u}'; t_0)$.  In other words, the matrix ${\bf A^{T}}(\vec{u},\vec{u}'; t)$ does not become less sparse through this process of correction (a fact which is important for large numerical matrix manipulations).

It is worth mentioning that whether updating the correction matrix is a viable method depends on the time scales of the changing ionosphere.  More specifically, the above was calculated keeping only terms to first order in $\Delta c_q/c_q(t_0)$.  It is possible that for the time scales considered higher terms are also necessary, or that $\Delta c_q$ is not small compared to $c_q(t_0)$; the former situation complicates the math but does not necessarily outrule this method, while the latter pretty much requires that the correction matrix be built from scratch every time.

\section{Conclusions}

\hspace{4 mm}  The $uv$-plane correction only makes computational sense if the model for the ionospheric perturbation allows for an analytic solution to $I(\vec{u})$ (equation \ref{IUpure}).  One such model is a sum over sinusoidal modes (equation \ref{dtsine}).  By running numerical codes with this choice, the most important result discovered was that under correcting in the $uv$-plane is worse than not correcting at all (section \ref{nmax}).  But in addition to this, correcting to too many orders or requiring too many modes to model the effect of the ionosphere may lead to a computationally unreasonable problem (section \ref{numterms}).  To help avoid this issue, a theoretical estimate of the number of orders of correction necessary (which agrees well with the sample sky provided in this paper) may be used (section \ref{theorynmax}).  This estimate reveals that the number of orders of correction necessary varies in the $uv$-plane.  This, however, suggests two methods for alleviating the problem:  eliminating those points in the $uv$-plane which are particularly troublesome at the cost of precision for the real space sky (section \ref{meth1}) and correcting to different orders at different points in the $uv$-plane (section \ref{meth2}).  Both techniques prove successful and make the problem of correcting in the $uv$-plane more feasible.  In addition, depending on how often the ionosphere's effect is updated compared to the timescales of change in the ionosphere, it may be compuationally favorable to update the previously determined effect of the ionosphere rather than rederive its full effect from scratch each time.

\section{Acknowledgments}
The authors would like to thank Matias Zaldarriaga for helpful conversations.

\section*{Appendix}
\setcounter{section}{0}
\setcounter{equation}{0}
\def\thesection{\Alph{section}}

The purpose of this appendix is to rigorously derive some of the mathematical formulas merely stated within the main text.  It is included for completeness and for the curious reader; no new results are derived.

\section{Solving for $I(\vec{u})$ for a Sinusoidal Reflection}
\label{math1}

In this section, we solve for the unperturbed intensity $I(\vec{u})$,
\begin{equation}
I(\vec{u}) = \sum_{n=0}^{\infty} { \int{ \frac{d^2u'}{(2\pi)^2} \left( \int{ d^2 \theta' e^{i\vec{\theta'} \cdot (\vec{u'}-\vec{u})} \frac{ (-i \vec{u} \cdot \vec{\delta \theta'})^{n}}{n!}    } \right) \tilde{I}(\vec{u'}) } }
\end{equation}
using a sum over sinusoidal modes for our ionospheric deflection $\vec{\delta \theta'}$, 
\begin{equation}
\vec{\delta \theta'} = Re( \sum_{m=1}^{M} ia_m \vec{b}_m e^{-i \vec{b}_m \cdot \vec{\theta'} }  ).
\end{equation}
With a bit of algebra (and keeping in mind that $\vec{b}_m$ and $\vec{\theta'}$ are real but $a_m$ is complex), $\vec{\delta \theta'}$ may be written as 
\begin{equation}
\vec{\delta \theta'} = \sum_{m=1}^{M} \left( i \frac{a_m \vec{b}_m}{2} e^{-i \vec{b}_m \cdot \vec{\theta'} } - i \frac{a_m^* \vec{b}_m}{2} e^{i \vec{b}_m \cdot \vec{\theta'} } \right) ,
\end{equation}
where $a_m^*$ is the complex conjugate of $a_m$.  With this choice of $\vec{\delta \theta'}$, the expression for $I(\vec{u})$ becomes
{\scriptsize
\begin{eqnarray}
I(\vec{u}) & = & \int \frac{d^2u'}{(2\pi)^2} \int d^2 \theta' e^{i \vec{\theta'} \cdot (\vec{u'}-\vec{u})}  \sum_{n=0}^{\infty} \frac{1}{2n!}  \\
& & \mbox{ {\textbf X }  }  (  \sum_{m=1}^{M}  a_m \vec{u} \cdot \vec{b}_m e^{-i \vec{b}_m \cdot \vec{\theta'} } - a_m^* \vec{u} \cdot \vec{b}_m e^{i \vec{b}_m \cdot \vec{\theta'} } )^{n}   \tilde{I}(\vec{u}') .
\end{eqnarray}
}

Before proceeding, it is convenient to convert the summation over $m$ as follows:
\begin{eqnarray}
\sum_{m=1}^{M} & & \left( \frac{a_m \vec{u} \cdot \vec{b}_m}{2} e^{-i \vec{b}_m \cdot \vec{\theta'} } - \frac{a_m^* \vec{u} \cdot \vec{b}_m}{2} e^{i \vec{b}_m \cdot \vec{\theta'} } \right) \\
& = & \sum_{q=1}^{2M} \left( c_q \vec{u} \cdot \vec{d}_q  e^{-i \vec{d}_q \cdot \vec{\theta'} } \right),
\end{eqnarray}
where
\begin{equation}
 c_q = \left\{ \begin{array}{ll}
         \frac{1}{2} a_q & \mbox{if $q <  M+1$};\\
         \frac{1}{2} a_{q-M}^* & \mbox{if $q \geq M+1 $}.\end{array} \right. 
\end{equation} 
and
\begin{equation}
 \vec{d}_q = \left\{ \begin{array}{ll}
         \vec{b}_q & \mbox{if $q < M+1$};\\
         - \vec{b}_{q-M} & \mbox{if $q \geq M+1$}.\end{array} \right. 
\end{equation} 
With this form, we see that although there are $M$ modes distorting the sky, there are $2M$ terms in the sum.  This extra factor of 2 comes from the above constraint that $\vec{\delta \theta'}$ be real.  We will refer to these modes labeled by ($  c_q, \vec{d}_q  $ )  as \emph{effective modes}.  Writing the intensity $I(\vec{u})$ in terms of effective modes gives
{\small
\begin{eqnarray}
I(\vec{u}) & = &  \int \frac{d^2u'}{(2\pi)^2} \left( \int d^2 \theta' e^{i\vec{\theta'}(\vec{u'}-\vec{u})} \right. \\ 
& & \mbox{ {\textbf X } } \left. \sum_{n=0}^{\infty} \frac{1}{n!} \left(  \sum_{q=1}^{2M}  c_q \vec{u} \cdot \vec{d}_q  e^{-i \vec{d}_q \cdot \vec{\theta'} }  \right)^{n}    \right) \tilde{I}(\vec{u}') .
\end{eqnarray}
} 
The individual terms inside the summation over $n$ may be manipulated using the multinomial expansion to give
{ \small 
\begin{eqnarray}
\frac{1}{n!} & & \left( \sum_{q=1}^{2M} c_q \vec{u} \cdot \vec{d}_q e^{-i \vec{d}_q \cdot \vec{\theta'} } \right)^{n} = \\
& = & \frac{1}{n!} \sum_{l_1, l_2, ... , l_{2M}}' \frac{n!}{l_1! l_2! ... l_{2M} !} \prod_{q=1}^{2M} \left( c_q \vec{u} \cdot \vec{d}_q e^{-i \vec{d}_q \cdot \vec{\theta'} } \right)^{l_q} \\
& = & \sum_{l_1, l_2, ... , l_{2M}}' e^{-i  \vec{\theta'} \cdot \sum_q \vec{d}_q l_q} \prod_{q=1}^{2M} \left( \frac{ (c_q \vec{u} \cdot \vec{d}_q)^{l_q}}{l_q!} \right) ,
\end{eqnarray}
}
where $\sum'$ denotes a restricted sum such that $\sum_q l_q = n$ and $l_q \geq 0$.  The expression for $I(\vec{u})$ now becomes
{ \scriptsize
\begin{eqnarray}
I(\vec{u}) & = & \int \frac{d^2u'}{(2\pi)^2} \left( \int d^2 \theta' e^{i\vec{\theta'}(\vec{u'}-\vec{u})}  \right. \\
& & \mbox{ {\textbf X} } \left. \sum_{n=0}^{\infty} \sum_{l_1, l_2, ... , l_{2M}}' e^{-i  \vec{\theta'} \cdot \sum_q \vec{d}_q l_q}  \prod_{q=1}^{2M} \frac{ (c_q \vec{u} \cdot \vec{d}_q)^{l_q}}{l_q!} \right) \tilde{I}(\vec{u'}) .
\end{eqnarray}
}
Because of the above performed multinomial expansion, the integral over $\vec{\theta'}$ (once brought inside the summation) now takes on the familiar form of a delta-function, and is easily performed to yield
{ \scriptsize
\begin{equation}
I(\vec{u}) = \int d^2 u' \sum_{n=0}^{\infty} \sum_{l_1, l_2, ... , l_{2M}}' \delta(\vec{u'} - \vec{u} - \sum_q \vec{d}_q l_q) \left( \prod_{q=1}^{2M} \frac{ (c_q \vec{u} \cdot \vec{d}_q)^{l_q}}{l_q!} \right) \tilde{I}(\vec{u'}) .
\end{equation}
}
The integral over $\vec{u'}$ is now a simple delta-function integral, and its integration gives
{ \footnotesize
\begin{equation}
\label{finalIu}
I(\vec{u}) = \sum_{n=0}^{\infty} \sum_{l_1, l_2, ... , l_{2M}}' \left( \prod_{q=1}^{2M} \frac{ (c_q \vec{u} \cdot \vec{d}_q)^{l_q}}{l_q!} \right) \tilde{I}(\vec{u} + \sum_q \vec{d}_q l_q) .
\end{equation}
}

\section{Finding an Upper Bound on the Error}
\label{math2}

\hspace{4 mm} In this section, we estimate the error in the $uv$-plane accumulated by truncating the infinite sum over $n$ after the term $n = n_{max} + 1$.  Recall that the total correction term is
{\small
\begin{equation}
I(\vec{u}) = \sum_{n=0}^{\infty} \sum_{l_1, l_2, ... , l_{2M}}' \left( \prod_{q=1}^{2M} \frac{ ( c_q \vec{u} \cdot \vec{d}_q)^{l_q}}{l_q!} \right) \tilde{I}(\vec{u} + \sum_q \vec{d}_q l_q) .
\end{equation}
} 
When correcting through $n=n_{max}$, the magnitude error $E(\vec{u})$ in the correction is equal to the absolute value of the sum over all terms left out.  More specifically, 
{\footnotesize
\begin{equation}
E(\vec{u}) = \left| \sum_{n=n_{max} + 1}^{\infty} \sum_{l_1, l_2, ... , l_{2M}}' \left( \prod_{q=1}^{2M} \frac{ ( c_q \vec{u} \cdot \vec{d}_q)^{l_q}}{l_q!} \right) \tilde{I}(\vec{u} + \sum_q \vec{d}_q l_q) \right| .
\end{equation}
} 
We now attempt to determine an upper bound $U(\vec{u})$ on this error.  To begin, we bring the absolute value inside the sum, so that all terms now add constructively, 
{\footnotesize
\begin{equation}
\sum_{n=n_{max} + 1}^{\infty} \sum_{l_1, l_2, ... , l_{2M}}' \left| \left( \prod_{q=1}^{2M} \frac{ ( c_q \vec{u} \cdot \vec{d}_q)^{l_q}}{l_q!} \right) \tilde{I}(\vec{u} + \sum_q \vec{d}_q l_q) \right| \geq E(\vec{u}) .
\end{equation}
} 
We expect the magnitude of $I(\vec{u})$ will be approximately the same at all points in the $uv$-plane.  Denoting the maximum value of $|\tilde{I}(\vec{u})|$ for the uncorrected sky as $I_{max}$, we find
\begin{equation}
\sum_{n=n_{max} + 1}^{\infty} \sum_{l_1, l_2, ... , l_{2M}}' I_{max} \left| \prod_{q=1}^{2M} \frac{ ( c_q \vec{u} \cdot \vec{d}_q)^{l_q}}{l_q!} \right| \geq E(\vec{u})
\end{equation}
Next, consider the terms in the sum of the form $c_q \vec{u} \cdot \vec{d}_q$.  Define $v(\vec{u})$ to be the maximum value of $| c_q \vec{u} \cdot \vec{d}_q |$ for a given value of $\vec{u}$ and the effective modes in question.  Then,
{ \footnotesize
\begin{eqnarray}
\sum_{l_1, l_2, ... , l_{2M}}' \left( \prod_{q=1}^{2M} \frac{ v(\vec{u}) ^{l_q}}{l_q!} \right) & = & v(\vec{u}) ^{n}\sum_{l_1, l_2, ... , l_{2M}}' \left( \prod_{q=1}^{2M} \frac{ 1 }{l_q!} \right) \\
& \geq & \sum_{l_1, l_2, ... , l_{2M}}' \left( \prod_{q=1}^{2M} \frac{ ( c_q \vec{u} \cdot \vec{d}_q)^{l_q}}{l_q!} \right) ,
\end{eqnarray}
} 
where the equality in the equation above occurs because the restricted sum over $l_q$ requires that $\sum l_q = n$.  With this substitution, the upper bound on our error function becomes
\begin{equation}
\sum_{n=n_{max} + 1}^{\infty} I_{max}   v(\vec{u}) ^{n} \sum_{l_1, l_2, ... , l_{2M}}' \left( \prod_{q=1}^{2M} \frac{ 1 }{l_q!} \right) \geq E(\vec{u}).
\end{equation}
Now concentrate on the inner summation.  Define $G$ according to 
\begin{equation}
G = max \left( \prod_{q=1}^{2M} \frac{ 1 }{l_q!} \right) \hbox{ \hspace{4 mm} where \hspace{4 mm}} \sum_{q=1}^{2M} l_q = n .
\end{equation}
I assert that
\begin{equation}
G = \frac{1}{\left( (\frac{n}{2M})! \right)^{2M} } .
\end{equation}
[Aside: Before providing the proof of this, we should point out that the fraction $n/(2M)$ is not guaranteed to be an integer, and therefore this factorial and the ones given hereafter should be taken to be given by the Gamma-function $\Gamma(z)$, 
\begin{equation}
z! = \Gamma(z+1) = \int_0^{\infty} t^z e^{-t} dt .]
\end{equation}
The proof is quite short:

I) Start with the given form of $G$, corresponding to $l_q = n/(2M)$, for any $q$.

II) Any value of $\prod \frac{ 1 }{l_q!}$ which corresponds to an alternative choice for the $l_q$ under the constraint that $\sum l_q = n$ may be obtained by multiplying this value for $G$ by a finite number of factors whose magnitudes are all less than 1. Therefore, $G$ is the maximum.  QED.

An example may be quite useful here: Consider the scenario with $n=12$ and $M=2$.  In this case, we assert that 
\begin{equation}
G = \frac{1}{3!3!3!3!},
\end{equation}
which corresponds to the choice $l_q = 3$ for all four $l_q$.  Now, let's pick an alternative choice for the $l_q$; let's say $l_1 = 5, l_2 = 2, l_3 = 4, l_4 = 1$.  For this choice we obtain,
\begin{equation}
G' = \frac{1}{5!2!4!1!}.
\end{equation}
But this may be re-written as 
\begin{equation}
G' = \frac{1}{3!3!3!3!}\left(\frac{3}{4}\right)\left(\frac{3*2}{4*5}\right) = G * (\hbox{ factors $< 1$})  ,
\end{equation}
where the first factor transforms $(l_2, l_3)$ from (3,3) to (2,4) and the second transforms $(l_1, l_4)$ from (3,3) to (5,1).  Therefore $G' < G$.

Our new knowledge of $G$, when combined with our previous determination of the number of terms in the restricted sum over $l_q$ (see section \ref{numterms}), leads us to conclude that
{\scriptsize
\begin{eqnarray}
  \mbox{ (\# terms) * max } \left( \prod_{q=1}^{2M} \frac{ 1 }{l_q!} \right) & = &  \frac{(n + 2M - 1)!}{n!(2M-1)!} \frac{1}{\left( (\frac{n}{2M})! \right)^{2M} } \\
& \geq &  \sum_{l_1, l_2, ... , l_{2M}}' \left( \prod_{q=1}^{2M} \frac{ 1 }{l_q!} \right) .
\end{eqnarray}
}
Therefore, our new upper bound $U(\vec{u})$ on the error $E(\vec{u})$ becomes
{\footnotesize
\begin{eqnarray}
\label{Uu}
U(\vec{u}) & = & \sum_{n=n_{max} + 1}^{\infty} I_{max}   v(\vec{u}) ^{n} \frac{(n + 2M - 1)!}{n!(2M-1)!} \frac{1}{\left( (\frac{n}{2M})! \right)^{2M} } \\
& \geq & E(\vec{u}).
\end{eqnarray}
}
This is our final result for a strict upper bound on the total error.  Notice that this final step is equivalent to assuming that the contributions from all modes are as strong as the strongest, and add constructively.  As such, $U(\vec{u})$ is clearly an upperbound on the error.  For one mode, this last step does not lead to that great of an overestimate.  With the addition of more modes, however, this step overemphasises the contribution from weaker modes, and leads to a (potentially much) larger overestimate of the error.

\section{Theoretical Prediction for $n_{max}$}
\label{math3}

\hspace{4 mm}  The goal of this section is to determine the value of $n_{max}$ necessary to obtain a precision in the $uv$-plane equal to $P$ if given $\vec{u}$ and the ionospheric modes distorting the sky.  Based upon the terrible consequences which result from undercorrecting in the $uv$-plane (see section \ref{nmax}), we begin with the expression just derived for the upperbound on the error in hopes of avoiding this pitfall. 

In order to make sense of our expression for an upper bound $U(\vec{u})$ (equation \ref{Uu}) and derive from it the optimal choice of $n_{max}$, we must make a few further approximations.  Some of these approximations will actually slightly decrease the expression for the error relative to $U(\vec{u})$, but are necessary in order to make sense of this ugly expression.

To begin, we define
\begin{equation}
g_n \equiv I_{max}   v(\vec{u}) ^{n} \frac{(n + 2M - 1)!}{n!(2M-1)!} \frac{1}{\left( (\frac{n}{2M})! \right)^{2M} }
\end{equation}
so that
\begin{equation}
U(\vec{u}) = \sum_{n = n_{max}+1}^{\infty} g_n .
\end{equation}
Next, we calculate the ratio $g_{n+1}/g_n$ and find that 
{\footnotesize
\begin{equation}
\frac{g_{n+1}}{g_n} = v(\vec{u}) \frac{(n+1+2M-1)}{(n+1)} \frac{(2M)}{(n+1)} \left( \frac{n}{n+1}\right)^M \left( \frac{e}{(1 + \frac{1}{n})^{n}} \right) ,
\end{equation}
}
where we have employed Stirling's Approximation, 
\begin{equation}
n! \approx (2 \pi n)^{1/2} n^n e^{-n} .
\end{equation}
Stirling's approximation is best suited for large $n$, but is actually quite accurate for small $n$ as well.  It gives an answer within 8\% of the actual value for $n=1$, and within 1\% for $n=9$.  In other words, by using this approximation we greatly simplify our expression and sacrifice only a little in terms of accuracy.  Using the fact that 
\begin{equation}
\lim_{n \rightarrow \infty} (1 + \frac{1}{n})^n = e ,
\end{equation}
we see that for large $n$ this ratio reduces to
\begin{equation}
\lim_{n \rightarrow \infty}  \frac{g_{n+1}}{g_n} = \frac{2Mv(\vec{u})}{n+1} .
\end{equation}
This expression shows that at a critical value of $n$, namely,
\begin{equation}
n_c = 2Mv(\vec{u}),
\end{equation}
this ratio is approximately equal to 1, and $g_{n_c + 1} \approx g_{n_c}$ .  For $n<n_c$, $g_{n+1} > g_n$; and for $n>n_c$, $g_{n+1} < g_n$.  In other words, $g_n$ is an increasing function of $n$ until $n_c$, and then decreases from then on. 

Strictly speaking, these results are only valid for large $n >> 2M$.  However, in order to get an approximate expression for $n_{max}$, we now extend these results to all $n$.  The hope is that the approximate value of $n_{max}$ is varied only slightly by this extension to small $n$.  But even if this approximation causes large enough error to raise doubts about our quantitative results for $n_{max}$, it should still be good enough to learn something about the qualitative behavior of $n_{max}$.  Recall that setting $n_{max} = 0$ is the same thing as not correcting in the $uv$-plane (see equation \ref{finalIU});
\begin{equation}
n_{max} = 0:  \hspace{5 mm} I(\vec{u}) = \tilde{I}(\vec{u}).
\end{equation}
Put differently, the $n=0$ term in the sum is of the order of the uncorrected $uv$-plane, $g_0 = \tilde{I}(\vec{u})$.  According to the above, successive corrections $g_n$ differ in magnitude from the previous term by a factor of $2Mv(\vec{u})/n$.  Therefore, the approximate magnitude of the term $g_n$ is
{ \footnotesize
\begin{eqnarray}
|g_n|  & \approx & |\tilde{I}(\vec{u})| \left(\frac{2Mv(\vec{u})}{1}\right) \left(\frac{2Mv(\vec{u})}{2}\right) ... \left(\frac{2Mv(\vec{u})}{n}\right) \\
& \approx & |\tilde{I}(\vec{u})| \frac{(2Mv(\vec{u}))^n}{n!} .
\end{eqnarray}
}

We expect the distortions created by the sky to alter the magnitude of the intensity only very slightly, so that $|\tilde{I}(\vec{u})| \approx |I_{actual}| $.  Furthermore, for $n > n_c$ (which is the case for $n_{max}$) we approximate that the ratio $g_{n+1}/g_n$ falls quick enough that we may approximate the total remaining error as being enirely due to $g_n$, $g_{n} \approx | I_{n} - I_{actual} |$, where $I_{n}$ is the intensity in the $uv$-plane after being corrected to $n$ orders. Therefore, in order to obtain an fractional error $f$ less than $P$ for our $uv$-plane correction, we must correct to enough orders $n$ so that    
\begin{equation}
P \geq f = \frac{| I_{n} - I_{actual} |}{|I_{actual}|} \approx \frac{(2Mv(\vec{u}))^n}{n!}. 
\end{equation}
Therefore the optimal value of $n_{max}$ is given by
\begin{equation}
n_{max} = \mbox{\hspace{1 mm} min \{  } n \mbox{ \} \hspace{1 mm} such that \hspace{1 mm} } 
Pn! \geq (2Mv(\vec{u}))^n .
\end{equation}
Some values for $n_{max}$ given $2Mv(\vec{u})$ and $P$ are given in the table embedded within the main text, table \ref{nmaxtab}.

\section{Deriving $I(\vec{u})$ for an Alternate choice for $\vec{\delta \theta}'$}
\label{math4}

\hspace{4 mm} A bulk of this paper has assumed that $\vec{\delta \theta}'$ takes the form

\begin{equation}
\delta \vec{\theta}' = Re( \sum_{m=1}^{M} ia_m \vec{b}_m e^{-i \vec{b}_m \cdot \vec{\theta}' }  ).
\end{equation}
This is, of course, only one of many possible choices.  In this section, we analyze the results of instead choosing

\begin{equation}
\delta \vec{\theta}' = \sum_{k=1}^{K} a_k \theta_x'^{k} + \sum_{m=1}^{M} b_m \theta_y'^{m} , 
\end{equation}
where $K$ and $M$ are the number of terms in the $\theta_x$ and $\theta_y$ directions, respectively, necessary to accurately model the distortion by the ionosphere.  The $a_k$ and $b_m$ here are real.  Similar to last time, this choice is chosen because it allows an analytic solution to $I(\vec{u})$.  With this choice, our earlier equation for $I(\vec{u})$,
\begin{equation}
I(\vec{u}) = \int \frac{d^2u'}{(2\pi)^2} \int d^2 \theta' e^{i \vec{\theta}' \cdot (\vec{u}' - \vec{u}) } e^{ -i \vec{u} \cdot \delta \vec{\theta}' } \tilde{I}(\vec{u}') , 
\end{equation}
becomes
{\footnotesize
\begin{eqnarray}
I(\vec{u}) & = & \int \frac{d^2u'}{(2\pi)^2} \int d^2 \theta' e^{i \vec{\theta}' \cdot (\vec{u}' - \vec{u}) }\left( e^{- i u_x \sum_k a_k \theta_x'^k } \right)  \\
& & \mbox{ {\textbf X} } \left(e^{-i u_y \sum_m b_m \theta_y^m } \right) \tilde{I}(\vec{u}') \\
& = & \int \frac{du_x'}{2\pi} \int d\theta_x' e^{i \theta_x' (u_x' - u_x) } \left(e^{-i u_x \sum_k a_k \theta_x^k } \right) I_y
\end{eqnarray}
}
where
\begin{equation}
I_y \equiv \int \frac{du_y'}{2\pi} \tilde{I}(\vec{u}') \int d\theta_y' e^{i \theta_y' (u_y' - u_y) } \left(e^{-i u_y \sum_m b_m \theta_y^m } \right)
\end{equation}
First, we focus on evaluating $I_y$.  To do this, we first Taylor-expand the second exponential,
{ \footnotesize
\begin{eqnarray}
e^{-i u_y \sum_m b_m \theta_y^m } & = & \sum_{n=0}^{\infty} \frac{1}{n!} \left( -iu_y \sum_{m=1}^{M} b_m \theta_y^m \right)^n \\
& = & \sum_{n=0}^{\infty} \frac{(-i u_y)^n}{n!} \sum_{l_m}' n! \prod_{m=1}^{M} \frac{(b_m \theta_y'^m)^{l_m}}{l_m!}
\end{eqnarray}
}
where the sum over $l_m$ is a restricted sum such that $\sum_m l_m = n$.  Plugging this into our expression for $I_y$ gives
{ \scriptsize
\begin{eqnarray}
I_y & = & \sum_{n=0}^{\infty} \sum_{l_m}' \int \frac{du_y'}{2\pi} \tilde{I}(\vec{u}') \int d\theta_y' e^{i \theta_y' (u_y' - u_y) } (-i u_y)^n \\
& & \mbox{ \hspace{3 mm} {\textbf X} } \prod_{m=1}^{M} \frac{(b_m \theta_y'^m)^{l_m}}{l_m!} \\
& = & \sum_{n=0}^{\infty} \sum_{l_m}' \int \frac{du_y'}{2\pi} \tilde{I}(\vec{u}') \int d\theta_y' e^{i \theta_y' (u_y' - u_y) } (-i u_y)^n \theta_y'^{\sum_m m l_m} \\
& & \mbox{ \hspace{3 mm} {\textbf X} } \prod_{m=1}^{M} \frac{(b_m)^{l_m}}{l_m!} ,\\
\end{eqnarray}
}
where we have assumed that the summations and the integrals may be freely interchanged.  To solve this integral, we first introduce an additional parameter $\lambda$ (which we will eventually set to 1) and notice that
{ \scriptsize
\begin{eqnarray}
\int d\theta_y' e^{i \theta_y' ( u_y' - \lambda u_y) } \theta_y'^t & = & \int d\theta_y' \frac{1}{(-i u_y)^t} \frac{\partial^t}{\partial \lambda^t} e^{ i \theta_y' (u_y' - \lambda u_y)} \\
& = & \frac{1}{(-i u_y)^t} \frac{\partial^t}{\partial \lambda^t} \int d\theta_y' e^{ i \theta_y' (u_y' - \lambda u_y)} \\
& = & \frac{2 \pi}{(-i u_y)^t} \frac{\partial^t}{\partial \lambda^t} \delta(u_y' - \lambda u_y)
\end{eqnarray}
}
Therefore, if we define $t = \sum_m m l_m$, then
{\small
\begin{eqnarray}
I_y & = & \sum_{n=0}^{\infty} \sum_{l_m}' \int du_y' \tilde{I}(\vec{u}') (-i u_y)^{n-t} \\ 
& & \mbox{ \hspace{3 mm} {\textbf X} } \prod_{m=1}^{M} \frac{(b_m)^{l_m}}{l_m!} \frac{\partial^t}{\partial \lambda^t} \delta(u_y' - \lambda u_y) \\
& = & \sum_{n=0}^{\infty} \sum_{l_m}' (-i u_y)^{n-t}  \prod_{m=1}^{M} \frac{(b_m)^{l_m}}{l_m!} \frac{\partial^t}{\partial \lambda^t} \tilde{I}(u_x, \lambda u_y) .
\end{eqnarray}
}
But,
\begin{equation}
\left. \frac{\partial^t}{\partial \lambda^t}  \tilde{I}(u_x, \lambda u_y) \right|_{\lambda = 1} = u_y \left( \frac{\partial^t \tilde{I}(\vec{u})}{\partial u_y^t} \right) .
\end{equation}
Thus,
{\small
\begin{equation}
I_y =  \sum_{n=0}^{\infty} \sum_{l_m}' (-i u_y)^{n-t}  \prod_{m=1}^{M} \frac{(b_m)^{l_m}}{l_m!} u_y \left( \frac{\partial^t \tilde{I}(\vec{u})}{\partial u_y^t} \right) .
\end{equation}
}
Plugging this into our earlier expression for $I(\vec{u})$, we obtain
\begin{equation}
I(\vec{u}) = \sum_{n=0}^{\infty} \sum_{l_m}' i (-i u_y)^{n-t+1} \prod_{m=1}^{M} \frac{(b_m)^{l_m}}{l_m!} \frac{\partial^t}{\partial u_y^t} I_x
\end{equation}
where
\begin{equation}
I_x = \int \frac{du_x'}{2\pi} \tilde{I}(\vec{u}') \int d\theta_x' e^{i \theta_x' (u_x' - u_x) } \left(e^{-i u_x \sum_k a_k \theta_x^k } \right) .
\end{equation}
But $I_x$ here is of the same form as $I_y$ earlier, and therefore
{\footnotesize
\begin{eqnarray}
I(\vec{u}) = & & \sum_{n_x}^{\infty} \sum_{n_y}^{\infty} \sum_{l_m}' \sum_{l_k}' - (-i u_x)^{n_x-t_x+1} (-i u_y)^{n_y-t_y+1} \\
& & \mbox{ {\textbf X} } \prod_{m=1}^{M} \frac{(b_m)^{l_m}}{l_m!} \prod_{k=1}^{K} \frac{(a_k)^{l_k}}{l_k!} \frac{\partial^{t_x}}{\partial u_x^{t_x}} \frac{\partial^{t_y}}{\partial u_y^{t_y}} \tilde{I}(\vec{u}) 
\end{eqnarray}
}
where
\begin{equation}
t_y = \sum_m m l_m , 
\end{equation}
\begin{equation}
t_x = \sum_k k l_k ,
\end{equation}
and the sums over $l_m$ and $l_k$ are restricted so that $\sum_k l_k = n_x$ and $\sum_m l_m = n_y$.

\end{document}